\documentclass[11pt,a4paper,fleqn]{article}
\usepackage[margin=2.6cm]{geometry}
\usepackage{etoolbox}
\usepackage{xspace}
\usepackage{xcolor}
\usepackage[hidelinks]{hyperref}
\usepackage{booktabs}

\usepackage[authoryear,longnamesfirst]{natbib}

\usepackage{float}
\usepackage{caption}
\usepackage{tabularx}

\usepackage[most]{tcolorbox}

\newtcblisting{codeblock}{
    colback=gray!8,
    colframe=gray!8,
    boxrule=0pt,
    arc=2pt,
    left=6pt,
    right=6pt,
    top=5pt,
    bottom=5pt,
    listing only,
    listing options={
        basicstyle=\ttfamily\small,
        breaklines=true
    }
}

\usepackage{algorithm}
\usepackage{algpseudocode}
\usepackage{amsmath}

\def\tsc#1{\csdef{#1}{\textsc{\lowercase{#1}}\xspace}}
\tsc{WGM}
\tsc{QE}

\begin{document}
\let\WriteBookmarks\relax
\def\floatpagepagefraction{1}
\def\textpagefraction{.001}

\title{Beyond Global Importance: A Heterogeneity Index for Sensitivity Analysis}
\author{Mariia Kozlova$^{1,*}$, Justus Helo$^{2}$, Pamphile T. Roy$^{3}$,\\
Julian Scott Yeomans$^{2}$, Andrea Saltelli$^{4}$}
\date{}
\maketitle
\begin{center}
\small
$^{1}$ LUT Business School, LUT University, Lappeenranta, Finland\\
$^{2}$ Schulich School of Business, York University, Toronto, Ontario, Canada\\
$^{3}$ Independent researcher, Vienna, Austria\\
$^{4}$ UPF Barcelona School of Management, Barcelona, Spain\\[0.4em]
$^{*}$ Corresponding author: \texttt{Mariia.Kozlova@lut.fi}
\end{center}

\begin{abstract}
Global sensitivity analysis quantifies how strongly inputs influence a model output, however, input importance often varies across the model space. We introduce a heterogeneity index \(H\) to quantify this variation. The index compares normalized regional sensitivity profiles using mean pairwise total-variation distance and its values range from zero to one. We test \(H\) on nine benchmark functions and examine its robustness to the number of regions, sample size, and sampling strategy. We further demonstrate it using models of wooden-pallet carbon footprint, flood risk, and steel-structure reliability. Across these cases, \(H\) reveals changes in sensitivity that global indices obscure, including inputs whose importance changes substantially across the model space and differences between rare behavioral regimes. The index complements global sensitivity analysis by indicating when a single global sensitivity profile is insufficient. An open-source implementation is provided in the \texttt{simdec} package.
\end{abstract}




\section{Introduction}

Environmental and energy models commonly represent complex systems shaped by nonlinear processes, interacting drivers, and interdependent uncertain conditions \citep{pianosi2016sensitivity,beven2001equifinality}. Consider a power-system model combining variable renewable generation, electricity demand, storage, and market or network constraints. A global sensitivity analysis may identify electricity demand as the dominant source of output variation overall, while under particular combinations of highly intermittent generation, high demand, or constrained import, other factors such as storage capacity or demand response flexibility may become substantially more influential. Such heterogeneity need not correspond to an explicit threshold or a regime known in advance; it may emerge only when the model is examined across different parts of its input or output space. This matters for decision-making because recommendations based on global importance can overlook the factors that become decisive under particular operating conditions. The term \textit{heterogeneity} is used in different ways across disciplines. In this study, we define heterogeneity as variation in the overall composition of input importance across the model space. 

Global sensitivity analysis (GSA) provides a rigorous framework for quantifying how uncertainty in model inputs contributes to uncertainty in model outputs and has become an established tool for model diagnosis, uncertainty assessment, and decision support \citep{tarantola2024annotated}. Most commonly used global sensitivity measures, however, summarize importance over the full distributions of the inputs and outputs. This aggregation can conceal variation in sensitivity within the model space. \cite{wei2013moment} noted that standard GSA measures input importance over its full range but does not identify how particular ranges contribute to output uncertainty, while  \cite{wei2015regional} subsequently developed regional extensions of Sobol' indices to examine how main and total effects change when input ranges are restricted. The same phenomenon has been demonstrated from a temporal perspective. Time-varying sensitivity analysis has shown that dominant model parameters can change substantially with environmental conditions and that these changes may be difficult to anticipate from the model formulation alone \citep{herman2013time}. Related work has therefore developed visualization approaches specifically for exploring changing sensitivity patterns across time and evaluation scales \citep{massmann2014new}. Collectively, these studies show that sensitivity itself can be heterogeneous: the factors that matter, and their relative importance, may change across the domain of a model.

Several branches of sensitivity analysis provide ways to investigate such variation. Regional sensitivity analysis partitions simulations into behavioral and non-behavioral subsets and examines differences between them \citep{spear1980eutrophication}, while later regional approaches investigate contributions from particular portions of the input domain \citep{wei2013moment,wei2015regional}. Target and conditional sensitivity analysis focus respectively on the factors driving entry into a critical output domain and on sensitivity within that domain \citep{marrel2021statistical}. For temporal or spatial outputs, cluster-based GSA identifies distinct model behaviors and the inputs associated with transitions between them \citep{roux2021cluster}. Graphical methods such as simulation decomposition (SimDec) provide another route by visualizing how conditional distributions and interactions change across combinations of model inputs, revealing heterogeneous effects that conventional global sensitivity indices may conceal \citep{Kozlova2024Heterogeneous}. These approaches differ substantially in their objectives and formulations, but share the recognition that a single global importance measure may provide an incomplete description of model behavior.

What remains difficult to express is the \emph{magnitude of heterogeneity itself}. Existing approaches can identify sensitivity within particular regions, trace changing indices across time or input ranges, distinguish critical and non-critical domains, or reveal heterogeneous patterns visually. They do not, however, provide a general scalar measure of how much the complete sensitivity profile changes across a partition of the model space. Consequently, there is no direct counterpart to a global sensitivity index for answering a simple diagnostic question: \emph{How heterogeneous is the sensitivity structure of this model?} Such a measure could complement conventional GSA by indicating when a global sensitivity profile is sufficiently representative and when, instead, more detailed regional investigation is warranted.

In this paper, we propose the heterogeneity index $H$ to quantify changes in sensitivity profiles across regions of a model. The general index $H_Z$ is defined from pairwise distances between normalized regional sensitivity profiles, where $Z$ denotes the variable used to partition the model space. Regions may be defined by the model output, yielding $H_Y$; by an input, yielding $H_{X_i}$; or by an externally specified categorical partition representing application-specific states or behaviors. We first examine the index's properties using controlled models in which we can systematically introduce heterogeneity. An existing workbench of test functions for sensitivity analysis \citep{azzini2022function} is then used to investigate the influence of regionalisation and sample size, and to examine the relationship between global input importance and heterogeneity. Finally, three practical cases test whether $H$ can compactly identify forms of heterogeneous model behaviour previously revealed only through detailed visualisation or regional sensitivity analysis. The method is implemented in the free and open-source \texttt{simdec} package \citep{roy2024simulation,simdec2026github} to facilitate its use alongside conventional sensitivity analysis.

\section{Proposed heterogeneity index}
\label{sec:heterogeneity-index}

Let $Y=f(\mathbf X)$ be a model with inputs
$\mathbf X=(X_1,\ldots,X_K)$, and let $P$ denote a generic variable used to
define regions of the model space. In a standard analysis, $P$ is either the
output $Y$, one of the model inputs $X_j$, or an external
application-specific partitioning variable $Z$. The proposed heterogeneity index
$H_P$ measures how much the model's sensitivity structure changes across
regions defined by $P$.

Two cases arise naturally in a standard analysis. When $P=Y$, the resulting
index $H_Y$ measures changes in sensitivity structure across the output
distribution. When $P=X_j$, the index $H_{X_j}$ measures changes in
sensitivity structure across the range of input $X_j$. An external
partitioning variable $Z$ extends the same analysis to application-specific
conditions or regimes, such as safe/failure states, operating modes, or policy
scenarios, yielding $H_Z$. 

For a continuous partitioning variable $P$, its range is divided into $q$
equal-frequency regions $R_1,\ldots,R_q$. A categorical $P$ is divided
according to its observed states.

\subsection{Regional sensitivity profiles}

Within each region $R_r$, we compute a vector of \emph{combined}
sensitivity indices \citep{Kozlova2025Binning},
\begin{equation}
    \mathbf S_r
    =
    (S_{1,r},\ldots,S_{K,r})^{\top}.
    \label{eq:regional-profile-raw}
\end{equation}

The combined indices allocate estimated second-order interaction effects
between the participating inputs according to the Shapley allocation
principle.\footnote{
The first-order Sobol' sensitivity index is defined as
\[
S_i^{(1)}
=
\frac{
\operatorname{Var}\!\left[
\operatorname{E}(Y\mid X_i)
\right]
}{
\operatorname{Var}(Y)
},
\]
while the second-order interaction effect between $X_i$ and $X_j$ is
\[
S_{ij}^{(2)}
=
\frac{
\operatorname{Var}\!\left[
\operatorname{E}(Y\mid X_i,X_j)
\right]
}{
\operatorname{Var}(Y)
}
-
S_i^{(1)}
-
S_j^{(1)}.
\]
}
\begin{equation}
    S^{\mathrm{comb}}_{i,r}
    =
    S_{i,r}^{(1)}
    +
    \frac{1}{2}
    \sum_{j\neq i}
    S_{ij,r}^{(2)}.
    \label{eq:combined-si}
\end{equation}

Unlike first-order indices alone, which omit interactions, and total-effect
indices, which count interactions repeatedly, the combined indices distribute
second-order interaction effects across the participating inputs. Their sum
therefore approaches one when the full output variance is explained
\citep{Helo2026Regional}, provided that interactions above second order can be
neglected \citep{Kozlova2025Binning}.

The set of inputs included in a regional sensitivity profile depends on the
partitioning variable. If a categorical input $X_j$ is used as $P$, it is
constant within each region and is therefore omitted from the regional
sensitivity calculation. A continuous input used as $P$ remains in the
analysis because it still varies within each region. When an external
partitioning variable $Z$ is used, it serves only to assign observations to
regions and is not introduced among the inputs subjected to sensitivity
analysis. For example, $Z$ may distinguish behavioral and non-behavioral
model realizations in the terminology of \citet{spear1980eutrophication}.

Regional sensitivity analysis requires the output to retain variation within
each region,
$\operatorname{Var}(Y\mid R_r)>0$. Consequently, $H_Y$ is undefined for a
categorical output because partitioning the output according to its categories
makes $Y$ constant within each region and regional sensitivity indices cannot
be estimated. Such a model can nevertheless be analyzed using $H_{X_j}$ or,
where relevant, an external partition $H_Z$.

\subsection{Normalization and heterogeneity}

The sum of regional sensitivity indices may differ from the global estimate.
This is particularly pronounced when conditioning on the output. Multiple
compensating input configurations may generate the same output region,
producing regional equifinality and a loss of regional explainability
\citep{Helo2026Regional}.

To make the heterogeneity index respond to changes in the
\emph{composition} of input importance rather than to changes in the total
regional sensitivity sum, each regional profile is normalized:
\begin{equation}
    \widetilde S_{i,r}
    =
    \frac{S_{i,r}}
    {\sum_{j\in\mathcal I_P}S_{j,r}},
    \qquad
    \sum_{i\in\mathcal I_P}\widetilde S_{i,r}=1,
    \label{eq:normalized-profile}
\end{equation}
where $\mathcal I_P$ denotes the set of inputs retained in the regional
sensitivity calculation for partitioning variable $P$.

Non-normalized regional sensitivity indices remain useful diagnostics and are
retained for interpretation and visualization, whereas the normalized
profiles are used to calculate $H_P$. The consequences of this distinction,
including the removal of output-conditioned equifinality from the
heterogeneity measure, are illustrated in
Section~\ref{sec:tests-controlled}.

For two regions $R_r$ and $R_s$, the difference between their normalized
sensitivity profiles is measured by the total variation distance,
\begin{equation}
    D_{rs}
    =
    \frac{1}{2}
    \sum_{i\in\mathcal I_P}
    \left|
    \widetilde S_{i,r}
    -
    \widetilde S_{i,s}
    \right|.
    \label{eq:pairwise-tv}
\end{equation}

Because the normalized profiles are nonnegative and sum to one,
$D_{rs}\in[0,1]$. The factor $1/2$ avoids double counting sensitivity that
moves from one input to another: a decrease in one component must be
accompanied by an increase elsewhere in the profile.

For $q$ regions, the heterogeneity index is the mean distance over all unique
pairs of regional sensitivity profiles,
\begin{equation}
    H_P
    =
    \frac{1}{\binom{q}{2}}
    \sum_{r<s}D_{rs}
    =
    \frac{1}{\binom{q}{2}}
    \sum_{r<s}
    \frac{1}{2}
    \sum_{i\in\mathcal I_P}
    \left|
    \widetilde S_{i,r}
    -
    \widetilde S_{i,s}
    \right|.
    \label{eq:heterogeneity-index}
\end{equation}

Thus $H_P\in[0,1]$. A value of zero means that all regions have the same
relative sensitivity profile. Larger values indicate progressively stronger
redistribution of importance across the regions. Appending an irrelevant
input with zero sensitivity in every region does not affect $H_P$, because it
contributes zero to every pairwise distance.

The complete procedure is summarized in
Algorithm~\ref{alg:heterogeneity}.

\begin{algorithm}
\caption{Heterogeneity analysis}
\label{alg:heterogeneity}

\begin{algorithmic}

\State \textbf{1.} Obtain an input--output dataset $(\mathbf X,Y)$

\State \textbf{2.} Set the partitioning variables to
$\mathcal P=\{Y,X_1,\ldots,X_K\}$ for a standard analysis, or
$\mathcal P=\{Z\}$ for an external application-specific partition

\For{each partitioning variable $P\in\mathcal P$}

    \State \textbf{3.} Define regions $R_1,\ldots,R_q$ from $P$

    \For{each region $R_r$}
        \State \textbf{4.} Compute the combined sensitivity profile $\mathbf S_r$
        using Eq.~\eqref{eq:combined-si}
        \State \textbf{5.} Normalize the profile using
        Eq.~\eqref{eq:normalized-profile}
    \EndFor

    \State \textbf{6.} Compute pairwise profile distances using
    Eq.~\eqref{eq:pairwise-tv}

    \State \textbf{7.} Compute $H_P$ using
    Eq.~\eqref{eq:heterogeneity-index}

    \State \textbf{8.} Optionally compute $C_{i,P}$ using
    Eq.~\eqref{eq:heterogeneity-contribution}

\EndFor

\State \textbf{9.} Return $H_P$ for each analyzed partitioning variable
$P\in\mathcal P$

\end{algorithmic}
\end{algorithm}

\subsection{Individual contributions to heterogeneity}

Although $H_P$ compares complete sensitivity profiles, its value can be
decomposed into contributions from individual inputs. For input $X_i$, define
\begin{equation}
    C_{i,P}
    =
    \frac{1}{\binom{q}{2}}
    \sum_{r<s}
    \frac{1}{2}
    \left|
    \widetilde S_{i,r}
    -
    \widetilde S_{i,s}
    \right|.
    \label{eq:heterogeneity-contribution}
\end{equation}

Because the order of the finite sums can be interchanged,
\begin{equation}
    H_P
    =
    \sum_{i\in\mathcal I_P} C_{i,P}.
    \label{eq:h-contribution-sum}
\end{equation}

Thus, $C_{i,P}$ is the contribution of input $X_i$ to the total profile
difference summarized by $H_P$. Individual contributions are optional
diagnostics and are useful when a numerical attribution of heterogeneity to
particular inputs is desired. Otherwise, the regional sensitivity profiles
themselves show which components of the sensitivity structure change across
regions.

\paragraph{Simple two-variable example.}
To build intuition for the meaning of the heterogeneity index, consider a
model with two inputs, $X_1$ and $X_2$, and a partitioning variable $P$ that
defines two regions, $R_1$ and $R_2$. Their normalized sensitivity profiles
are
\begin{equation}
\widetilde{\mathbf S}
=
\begin{array}{c|cc}
    & R_1 & R_2 \\
    \hline
    X_1 & 0.8 & 0.2 \\
    X_2 & 0.2 & 0.8
\end{array}.
\label{eq:two-region-example}
\end{equation}

Because there is only one pair of regions,
\begin{equation}
    H_P
    =
    \frac{1}{2}
    \left(
    |0.8-0.2|
    +
    |0.2-0.8|
    \right)
    =
    0.6.
\end{equation}

The corresponding individual contributions are
\begin{equation}
    C_{1,P}=C_{2,P}=0.3,
\end{equation}
so that
\[
H_P=C_{1,P}+C_{2,P}=0.6.
\]

The interpretation is straightforward: $60\%$ of the sensitivity mass must be
reallocated to transform the sensitivity profile in one region into that in
the other. Adding a third input with zero sensitivity in both regions leaves
$H_P=0.6$, illustrating that the index is invariant to the addition of
irrelevant dimensions.

\subsection{Implementation}

The heterogeneity analysis is implemented in the open-source \texttt{simdec}
Python package. The package can be installed from PyPI using \texttt{pip install simdec} and imported in Python as follows:

\begin{codeblock}
import simdec
\end{codeblock}

For a standard analysis, the user provides the model inputs and
output:

\begin{codeblock}
H = simdec.heterogeneity_indices(
    output=y,
    inputs=X
)
\end{codeblock}

By default, continuous partitioning variables are divided into five
equal-frequency regions. The function calculates heterogeneity with respect to
the output and all model inputs simultaneously,

\[
H_Y,\;H_{X_1},\ldots,H_{X_K}.
\]

The number of regions can be changed using \texttt{n\_regions}:

\begin{codeblock}
H = simdec.heterogeneity_indices(
    output=y,
    inputs=X,
    n_regions=10
)
\end{codeblock}

An external application-specific partitioning variable $Z$ can be supplied
directly:

\begin{codeblock}
H = simdec.heterogeneity_indices(
    output=y,
    inputs=X,
    custom_partition=Z
)
\end{codeblock}

In this case, only $H_Z$ is calculated and the partitioning variable is not
introduced among the model inputs. If \texttt{custom\_partition} is
continuous, it is divided into \texttt{n\_regions} equal-frequency regions.
If it is categorical, its natural categories are used and
\texttt{n\_regions} is ignored.

The resulting heterogeneity indices are available through

\begin{codeblock}
H.indices
\end{codeblock}

while detailed results for each partition are stored in \texttt{H.details}.
For example,

\begin{codeblock}
H.details["X1"].raw_profiles
H.details["X1"].normalized_profiles
H.details["X1"].regional_sums
H.details["X1"].region_counts
H.details["X1"].individual_contributions
\end{codeblock}

The raw regional profiles contain the regional combined sensitivity indices,
whereas the normalized profiles are those used in the calculation of $H$.
Regional sensitivity-index sums and observation counts are retained as
diagnostics. Individual contributions $C_{i,P}$ are also returned for each
analyzed partitioning variable $P$ and sum to the corresponding $H_P$.

Regional sensitivity profiles can be visualized directly from the stored
results without recalculating the analysis:

\begin{codeblock}
H.plot()                  # all partitions
H.plot("X1")              # one partition
H.plot(["Y", "X1", "X2"]) # selected partitions
\end{codeblock}

The plots display the raw regional combined sensitivity indices, while the
heterogeneity index itself is calculated from normalized profiles.

Categorical variables are handled automatically. When a categorical input
$X_i$ defines the regions for $H_{X_i}$, that input is excluded from its own
regional sensitivity profiles because it is constant within each region. If
the model output is categorical, $H_Y$ is not computed because conditioning on
the output categories makes the output constant within each region. The
function issues an explanatory warning and continues to calculate
$H_{X_i}$ for the model inputs.

\section{Results: Heterogeneity versus global sensitivity}

Having defined the heterogeneity index, we first examine whether it captures information distinct from conventional global sensitivity analysis. We evaluate the global combined sensitivity indices $S_i$, input-conditioned heterogeneity $H_{X_i}$, and output-conditioned heterogeneity $H_Y$ across the benchmark functions summarized in Table~\ref{tab:benchmark_functions}. With the exception of the Binary model, which was constructed specifically for this study to provide a controlled case of regime-dependent sensitivity, the test functions are drawn from the benchmark collection of \citet{azzini2022function}. Figure~\ref{fig:h6} provides the central comparison.

Figure~\ref{fig:h6}(a) shows that global sensitivity and input-conditioned heterogeneity describe distinctly different properties of model behavior. Inputs occur throughout the resulting $S_i$--$H_{X_i}$ space: high global sensitivity may coincide with either high or low heterogeneity, while inputs with modest global sensitivity can exhibit substantial changes in the sensitivity structure across their ranges.

The contrast is particularly clear for F1 and F2. In both functions, $X_1$ has substantial global importance but essentially zero input-conditioned heterogeneity. Its contribution to output variance is therefore important, but the sensitivity structure remains stable as $X_1$ changes. Global importance alone does not imply heterogeneous model behavior.

\begin{center}
    \includegraphics[width=0.95\textwidth]{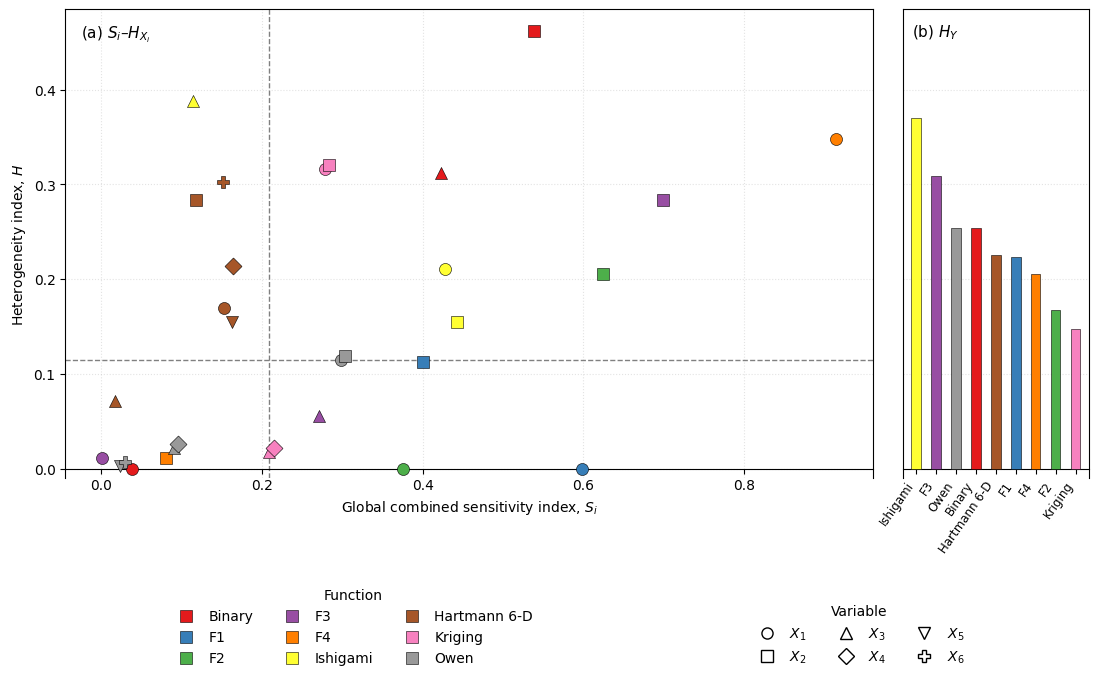}
    \captionof{figure}{
Relationship between global sensitivity and heterogeneity across the benchmark functions: 
(a) Global combined sensitivity index $S_i$ versus input-conditioned heterogeneity $H_{X_i}$;  
(b) output-conditioned heterogeneity $H_Y$.
        }
    \label{fig:h6}
\end{center}

\begin{table}[ht]
\centering
\caption{Test functions used to examine the sensitivity of $H$ to the number of regions.}
\label{tab:benchmark_functions}
\small
\renewcommand{\arraystretch}{1.35}

\begin{tabularx}{\textwidth}{
    >{\raggedright\arraybackslash}p{2.0cm}
    >{\raggedright\arraybackslash}X
    >{\raggedright\arraybackslash}p{3.2cm}
}
\hline
\textbf{Function} &
\textbf{Formulation} &
\textbf{Inputs and parameters} \\
\hline

Controlled binary &
$Y=X_1+(1+aX_2)X_3$ &
$X_1,X_3\sim U(0,1)$; $X_2\sim\mathrm{Bernoulli}(0.5)$; $a=4$ \\

F1 &
$f(\mathbf{x})=10x_1+0.5x_2^3$ &
$x_1,x_2\sim U(-5,5)$ \\

F2 &
$f(\mathbf{x})=2x_1-x_2^2$ &
$x_1,x_2\sim U(-5,5)$ \\

F3 &
$f(\mathbf{x})=
x_1^2+x_2^4+x_1x_2+x_2x_3^4$ &
$x_1,x_2,x_3\sim U(-4,4)$ \\

F4 &
$\displaystyle
f(\mathbf{x})=
0.2e^{x_1-3+2.2|x_2|}
+1.3x_2^6-2x_2^2
-0.5x_2^4-0.5x_1^4
+2.5x_1^2+0.7x_1^3
+\frac{3}{(8x_1-2)^2+(5x_2-3)^2+1}
+\sin(5x_1)\cos(3x_1^2)$ &
$x_1,x_2\sim U(-1,1)$ \\

Ishigami &
$f(\mathbf{x})=
\sin x_1+7\sin^2x_2+0.1x_3^4\sin x_1$ &
$x_1,x_2,x_3\sim U(-\pi,\pi)$ \\

Hartmann 6-D &
$\displaystyle
f(\mathbf{x})=
-\sum_{i=1}^{4}
\alpha_i
\exp\!\left[
-\sum_{k=1}^{6}
A_{ik}(x_k-P_{ik})^2
\right]$ &
$x_k\sim U(0,1)$; standard Hartmann 6-D parameters $\boldsymbol{\alpha},A,P$ \\

Kriging &
$\displaystyle
f(\mathbf{x})=
1+
e^{-2[(x_1-1)^2+x_2^2]-0.5(x_3^2+x_4^2)}
+
e^{-2[x_1^2+(x_2-1)^2]-0.5(x_3^2+x_4^2)}
$ &
$x_1,\ldots,x_4\sim U(0,1)$ \\

Owen &
$\displaystyle
f(\mathbf{x})=
\prod_{j=1}^{6}
\left[
\mu_j+\tau_j\sqrt{12}(x_j-0.5)
\right]$ &
$x_j\sim U(0,1)$; $\mu_j=1$; $\boldsymbol{\tau}=(1,1,0.5,0.5,0.25,0.25)$ \\

\hline
\end{tabularx}
\end{table}

The opposite pattern is visible in the Hartmann 6-D function. Five of its six inputs lie in the region of relatively low global sensitivity but above-median heterogeneity. Individually, these variables account for only modest portions of global output variance, yet conditioning on their values reveals pronounced changes in the model's sensitivity profile. This low-$S_i$, high-$H_{X_i}$ combination is particularly relevant for sensitivity analysis because such variables may receive little attention in a conventional global ranking despite being associated with substantially different local model behavior.

The Binary model provides the most extreme input-conditioned heterogeneity in the benchmark set. Its binary variable $X_2$ was deliberately constructed to switch the model between different sensitivity structures, and consequently produces the largest observed $H_{X_i}$. At the other end of the spectrum, the Owen function provides a multi-input example in which input-conditioned heterogeneity remains uniformly low: despite having six inputs with different levels of global importance, none produces a large change in the sensitivity profile across its range.

More generally, there is no common pattern in how heterogeneity is distributed among the inputs of a function. In some models, such as the Binary and Ishigami functions, substantial heterogeneity is concentrated in one particular input. In Hartmann 6-D it is distributed across most inputs, whereas in Owen it remains low throughout. Heterogeneity is therefore not simply a model-wide characteristic inherited equally by its inputs, but can itself have a highly heterogeneous structure.

Taken together, these results establish a distinctive property of the proposed heterogeneity index that sets it apart from conventional sensitivity indices. The global sensitivity index $S_i$ measures how much an input contributes to output variation overall. In contrast, $H_P$ measures how much the sensitivity structure changes across regions defined by the partitioning variable $P$. Consequently, high and low values of $S_i$ can coexist with either high or low heterogeneity. Sensitivity and heterogeneity therefore describe different dimensions of model behavior: $S_i$ identifies \emph{what matters}, whereas $H$ identifies \emph{how much what matters changes}.

\section{Validation and robustness}
\label{sec:heterogeneity-tests}

We evaluate the proposed index in three steps. First, controlled models are used to verify that $H$ responds to deliberately introduced heterogeneity and to illustrate the effect of normalizing the regional sensitivity profiles. Second, we test sensitivity to the number of regions. Third, we evaluate sample-size stability and repeatability under simple random and quasi-Monte Carlo sampling. 

\subsection{Raw versus normalized sensitivity indices}
\label{sec:tests-controlled}

To test whether the heterogeneity index behaves as intended, we use two simple controlled models in which heterogeneity is introduced through a single parameter $a$. At $a=0$, both models reduce to the additive form $A+B$, with a stable sensitivity structure across the model space. Increasing $a$ progressively introduces variation in the relative importance of the inputs, providing a controlled setting in which the response of $H$ can be examined.

All experiments use a scrambled Sobol' sample of $N=2^{17}=131{,}072$ observations. The parameter is varied over
$a\in\{0,0.25,0.5,0.75,1,2,4\}$.

The first model contains a binary regime variable,
\begin{equation}
    Y=A+(1+aK)B,
    \qquad
    A,B\sim U(0,1),
    \qquad
    K\sim\mathrm{Bernoulli}(0.5).
    \label{eq:binary-controlled}
\end{equation}

At $a=0$, the model reduces to $Y=A+B$ and the state of $K$ does not alter the sensitivity structure. As $a$ increases, the contribution of $B$ differs increasingly between the two states of $K$. We compute $H_K$ directly across the two states of $K$, and $H_Y$ across five equal-frequency regions of the output.

For illustration, Figure~\ref{fig:h1} shows heterogeneity calculated from both the raw regional sensitivity profiles and the normalized regional sensitivity profiles used in the proposed index. For input conditioning, the raw and normalized distances match when the raw regional sums are already close to one. For output conditioning, the distinction is more consequential. Even at $a=0$, output conditioning produces different raw regional sums because of equifinality \citep{Helo2026Regional}, so the raw profile distance exceeds 0 even though the underlying sensitivity composition is unchanged. Normalization removes this component and makes $H_Y$ reflect only changes in relative importance. 

The behavior of \(H_Y\) is not expected to increase smoothly with the heterogeneity control parameter \(a\). As \(a\) increases, \(H_K\) tracks the regime difference deliberately introduced through the binary input \(K\), and therefore changes monotonically with the strength of that mechanism. By contrast, \(H_Y\) measures heterogeneity after those input regimes have been mapped into the output space. Its value therefore depends not only on how different the \(K=0\) and \(K=1\) regimes are, but also on how their resulting output distributions overlap, separate, and mix within the chosen output regions. Consequently, stronger input-space heterogeneity does not necessarily translate into a proportionally larger \(H_Y\). At some values of \(a\), different input regimes may still populate the same output regions, while at others they become more clearly separated. Output-conditioned heterogeneity therefore reflects the combined effect of regime difference and the geometry of the input-to-output mapping.

\begin{center}
    \centering
    \includegraphics[width=0.95\textwidth]{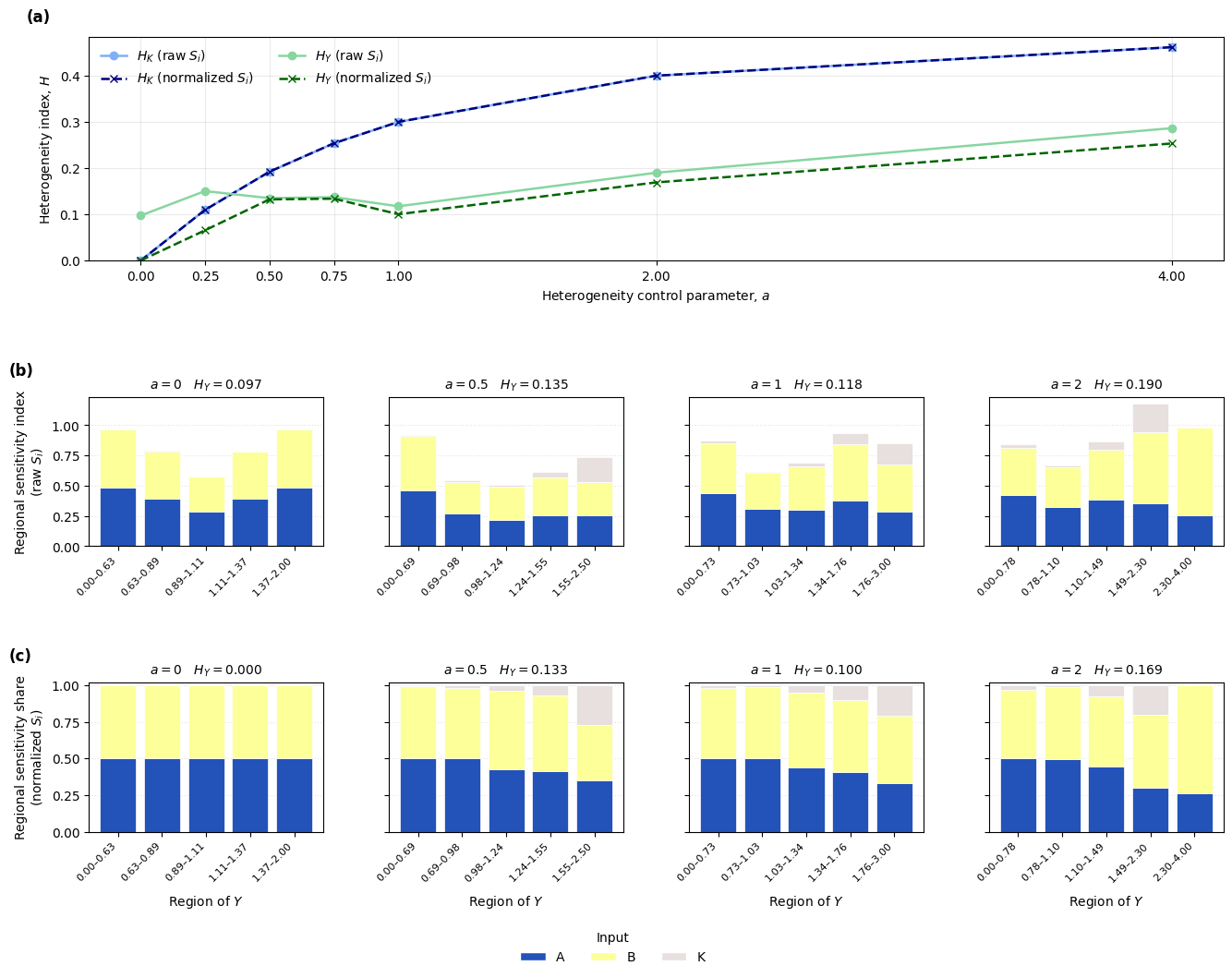}
    \captionof{figure}{Controlled binary model: (a) heterogeneity indices for \(K\) and \(Y\), computed using both raw and normalized regional sensitivity indices; (b) \(Y\)-conditioned regional raw sensitivity profiles, revealing a decline in regional explainability; and (c) \(Y\)-conditioned regional normalized sensitivity profiles, isolating changes in the relative sensitivity structure.}
    \label{fig:h1}
\end{center}

The second model replaces the binary regime with a continuous one,
\begin{equation}
    Y=A+(1+aC)B,
    \qquad
    A,B,C\sim U(0,1).
    \label{eq:continuous-controlled}
\end{equation}
The variable $C$ and the output $Y$ are each divided into five equal-frequency regions. At $a=0$, the model again reduces to $Y=A+B$. Increasing $a$ progressively changes the contribution of $B$ across the range of $C$. Figure~\ref{fig:h2} repeats the raw-versus-normalized comparison for the continuous model. The experiment verifies that the index can capture progressively changing, rather than only discrete, regimes. It also reinforces the distinction between input-space heterogeneity, $H_C$, and output-space heterogeneity, $H_Y$: the former measures the imposed structural change directly across $C$, whereas the latter measures how that change appears after projection into the output space.

\begin{center}
    \includegraphics[width=0.95\textwidth]{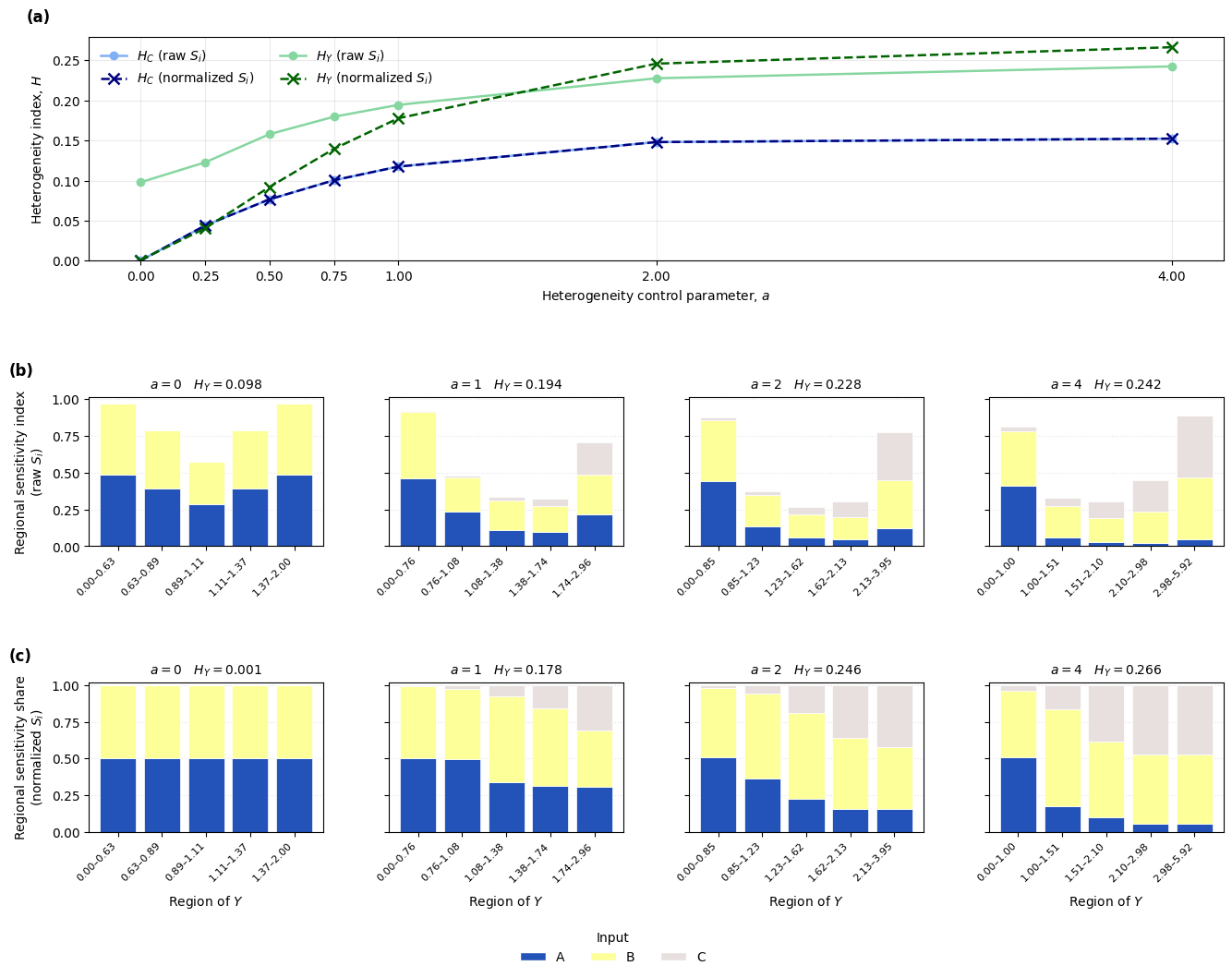}

    \captionof{figure}{
    Controlled continuous model: (a) heterogeneity indices for $C$ and $Y$, computed using both raw and normalized regional sensitivity indices; (b) $Y$-conditioned regional raw sensitivity profiles, showing changes in both sensitivity composition and regional explainability; and (c) $Y$-conditioned regional normalized sensitivity profiles, isolating changes in relative sensitivity structure.
    }
    \label{fig:h2}
\end{center}

\subsection{Sensitivity to the number of regions}
\label{sec:tests-regions}

The second experiment examines the dependence of $H$ on the number of regions used to partition a continuous conditioning variable. We use the controlled binary model in Eq.~\eqref{eq:binary-controlled} at $a=4$ and eight benchmark models \citep{azzini2022function}: F1, F2, F3, F4, Ishigami, Hartmann 6-D, Kriging, and Owen (Table \ref{tab:benchmark_functions}).

A single scrambled Sobol' sample of $N=2^{17}=131{,}072$ observations is generated for each model and reused across all partitions. For continuous conditioning variables, the number of equal-frequency regions is varied from $q=2$ to $q=15$; the binary variable retains its two natural states. The analysis is repeated for the output and for every model input.

Figure~\ref{fig:h3} shows that the estimated heterogeneity index can be sensitive to the number of regions, particularly when \(q\) is very small. Moving from \(q=2\) to \(q=3\) often produces pronounced jumps, whereas from approximately \(q=4\) onward the trajectories are more frequently smooth. In many cases, further increasing \(q\) gradually reduces the estimated heterogeneity, suggesting that finer partitioning can dilute broad differences in sensitivity structure across regions.

\begin{center}
    \includegraphics[width=0.95\textwidth]{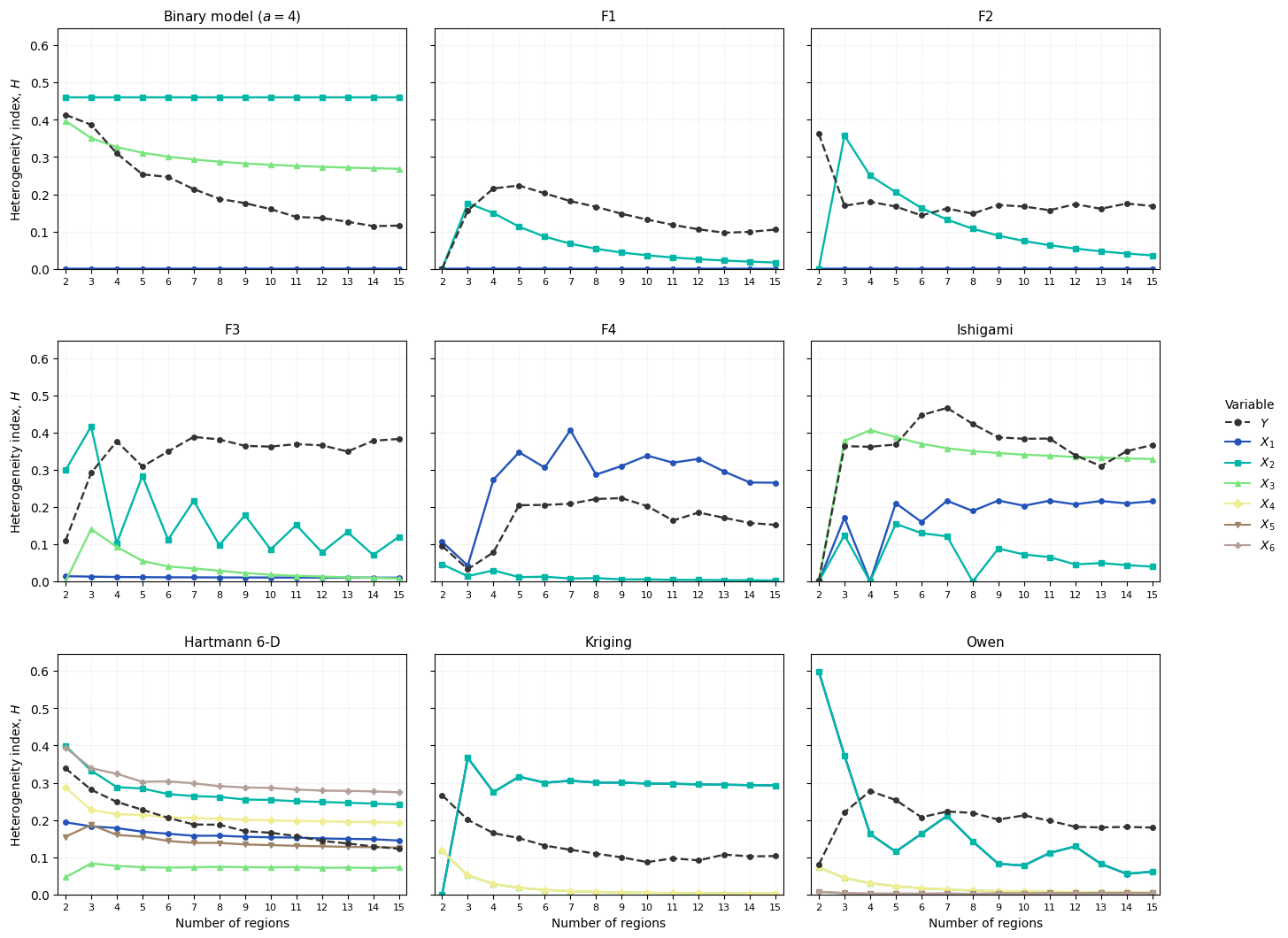}

    \captionof{figure}{
    Sensitivity of the heterogeneity index $H$ to the number of regions across the nine test models. Each panel shows $H$ obtained by conditioning on the output $Y$ and on each model input.
    }
    \label{fig:h3}
\end{center}

Several functions nevertheless show persistent non-monotonic behavior. F3 is the clearest example, with a pronounced odd-even pattern in \(q\). This behavior can be understood from its regional sensitivity profiles: a distinct sensitivity structure near the center of the output space is captured within a single region when \(q\) is odd, but split across two neighboring regions when \(q\) is even, reducing its contribution to the pairwise profile differences and hence to \(H\). Oscillatory functions, such as Ishigami, likewise show some non-monotonic dependence of \(H\) on \(q\), reflecting the interaction between the imposed regional partition and repeated local changes in model behavior. Appendix 1 presents the regional sensitivity profiles for all nine functions across several values of \(q\).

Overall, the experiment suggests that \(q=5\) is a reasonable starting point for estimating \(H\): it avoids the coarse-partition effects frequently observed for \(q=2\) or \(3\), while retaining sufficiently broad regions for stable regional sensitivity estimation. However, for black-box models, oscillatory responses, or models suspected to contain localized regime switches, heterogeneity should be examined across several values of \(q\) rather than relying on a single partition.

\subsection{Sample-size stability}
\label{sec:tests-sample-size}

We next fix the regional resolution at $q=5$ and investigate sensitivity to sample size and sampling strategy. The same nine models are evaluated using both simple random sampling (MC) and independently scrambled Sobol' quasi-Monte Carlo sampling (QMC). For MC, the sample sizes are $N=100$, $1{,}000$, $10{,}000$, and $100{,}000$. QMC uses the nearest powers of two, $N=128$, $1{,}024$, $8{,}192$, and $131{,}072$. Each model--method--sample-size combination is repeated ten times with an independently generated sample. Within each repetition, the same dataset is reused for all conditioning variables, and the quantile boundaries are re-estimated from that dataset.

The resulting distributions of $H$ reveal both the location and repeatability of the estimates as $N$ increases. Figure~\ref{fig:h4} shows the Kriging model as a representative example, with the MC and QMC results displayed side by side. Equivalent convergence plots for all nine models are provided in Appendix 2.

\begin{center}
    \includegraphics[width=0.95\textwidth]{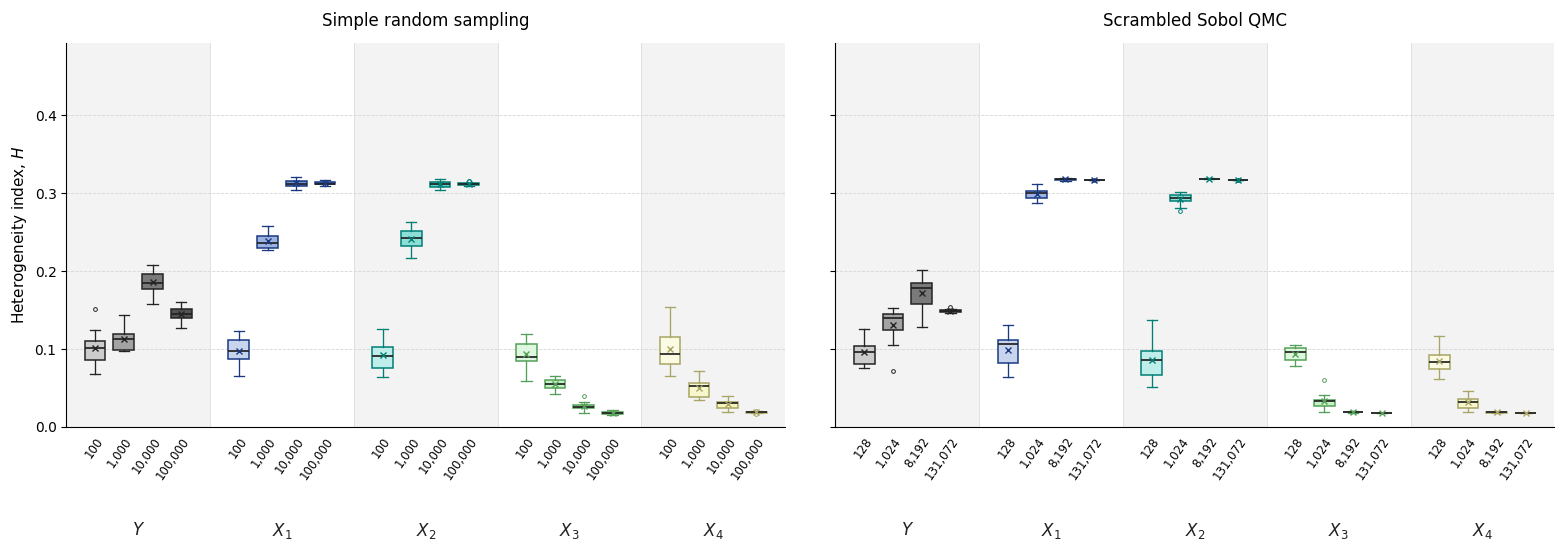}

    \captionof{figure}{
    Sample-size stability of $H$ for the Kriging function under Monte Carlo (MC) (left) and Sobol' quasi-Monte Carlo (QMC) sampling (right).
    }
    \label{fig:h4}
\end{center}

Figure~\ref{fig:h4} illustrates the effect of sample size for the Kriging function. At \(N=10^2\), estimates of \(H\) are clearly unreliable, with substantial dispersion across repetitions and pronounced bias relative to the larger-sample results. With \(q=5\), this sample size leaves only about 20 observations per region for estimating regional sensitivity indices, including second-order effects, so this instability is unsurprising. At \(N=10^3\), the estimates improve considerably, but noticeable bias and variability remain. From approximately \(N=10^4\) onward, the estimates become both stable and mutually consistent across sampling schemes. Small samples do not bias \(H\) in a single direction: for the Kriging model they underestimate heterogeneity for \(Y\), \(X_1\), and \(X_2\), while overestimating it for \(X_3\) and \(X_4\). There is no striking difference between simple random sampling and scrambled Sobol' QMC, except that the \(N\approx10^3\) estimates show a somewhat larger displacement between the two schemes.

\begin{center}
    \includegraphics[width=0.95\textwidth]{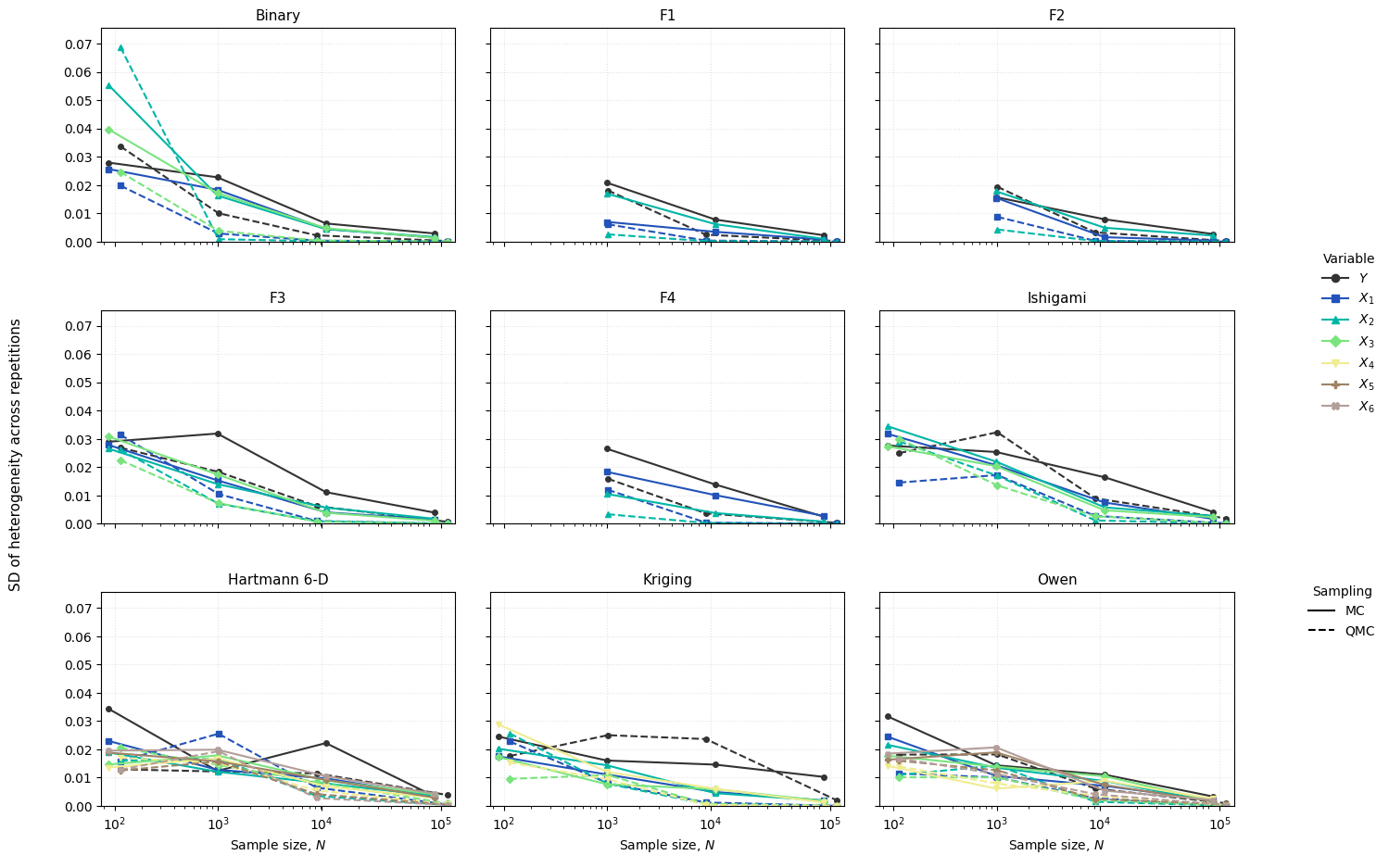}

    \captionof{figure}{
    Sample size and strategy effects on the heterogeneity index $H$ across all nine test functions.
    }
    \label{fig:h5}
\end{center}

To summarize repeatability across all models, Figure~\ref{fig:h5} reports the sample standard deviation of $H$ across the ten repetitions. The vertical axis therefore represents estimator variability. Actual sample sizes are used on the horizontal axis, and MC and QMC are shown separately as solid versus dashed line. At the smallest sample sizes, regional estimation can become degenerate when the internal sensitivity-estimation bins approach singleton occupancy.

Across all nine functions, the dispersion of \(H\) across repetitions generally decreases rapidly with sample size (Figure~\ref{fig:h5}). At the smallest usable samples, standard deviations are often on the order of \(0.02\)–\(0.07\), whereas by \(N\approx10^4\) they are typically below \(0.01\) and approach zero at \(N\approx10^5\). The pattern is broadly consistent across output- and input-conditioned estimates and across MC and QMC sampling, with neither sampling scheme showing a systematic precision advantage. Some functions, notably F3, Ishigami, Hartmann 6-D, and Kriging, retain somewhat greater variability for particular conditioning variables at intermediate sample sizes, reflecting more complex regional sensitivity structures. Overall, the results support \(N\approx10^4\) as a practical starting point for stable estimation in these benchmark models, while emphasizing that required sample size depends on both model structure and the number of observations retained within each region.

\section{Environmental case studies}
\label{sec:cases}

The practical contributions of the heterogeneity index are validated using three previously studied models in which heterogeneous behavior has already been identified visually. The first case revisits the carbon-footprint model studied by Kozlova et al.~\citep{Kozlova2024Heterogeneous}, where heterogeneous effects were previously established through simulation decomposition. The second considers a flood model examined extensively using regional sensitivity analysis \citep{Helo2026Regional}, allowing us to assess how well the heterogeneity index summarizes known variation in regional sensitivity profiles. The third revisits a steel structure reliability model exhibiting complex nested heterogeneous effects previously revealed through simulation decomposition \citep{ahola2024capturing}.

\subsection{Carbon footprint of wooden pallets}
\label{sec:case_wood}

The first case concerns the life-cycle carbon footprint of wooden transportation pallets. The model was originally introduced for simulation decomposition by Deviatkin et al. ~\citep{deviatkin2021simulation} and subsequently used by Kozlova et al.~\citep{Kozlova2024Heterogeneous} as an example of heterogeneous effects in environmental models. Total life-cycle emissions are calculated as

\[
E_{\mathrm{total}}
=
\left(
E_{\mathrm{production}}
+
E_{\mathrm{repair}}
+
E_{\mathrm{end-of-life}}
\right)
\frac{F}{N}
+
E_{\mathrm{transport}}F,
\]

where $F$ is the number of customer trips and $N$ is the number of uses of an individual pallet. Eight uncertain inputs describe the number of uses, transportation distance, materials and consumption, truck type, and end-of-life treatment. End of life is represented by two alternatives, incineration and landfilling (Table \ref{tab:wood_inputs}). Further details of the model can be found in \cite{Kozlova2024Heterogeneous}.

\begin{table}[ht]
\centering
\caption{Model inputs and uncertainty assumptions for the wooden-pallet carbon-footprint case.}
\label{tab:wood_inputs}
\small
\renewcommand{\arraystretch}{1.25}

\begin{tabularx}{\textwidth}{
    >{\raggedright\arraybackslash}p{3.2cm}
    >{\raggedright\arraybackslash}p{3.0cm}
    >{\raggedright\arraybackslash}X
    >{\raggedright\arraybackslash}p{1.5cm}
}
\toprule
\textbf{Input} &
\textbf{Distribution} &
\textbf{Parameters} &
\textbf{Unit} \\
\midrule

Timber &
Lognormal &
mean = 21.9, sd = 6.9 &
kg \\

Nails &
Lognormal &
mean = 0.35, sd = 0.11 &
kg \\

Electricity &
Lognormal &
mean = 0.69, sd = 0.73 &
kWh \\

Thermal energy &
Lognormal &
mean = 4.1, sd = 2.5 &
MJ \\

Number of uses &
Uniform &
1 to 83 &
uses \\

Transportation distance &
Uniform &
100 to 200 &
km \\

Truck type &
Discrete uniform &
big / medium / small, each $p=1/3$ &
-- \\

End of life &
Discrete uniform &
incineration / landfilling, each $p=1/2$ &
-- \\

\bottomrule
\end{tabularx}
\end{table}

The previous analysis was based on 1,000 Monte Carlo simulations. Global sensitivity analysis identified end of life and the number of uses as the two dominant inputs with a substantial interaction between them. Simulation decomposition then revealed a particular heterogeneous effect: the direction of the effect of the number of uses depends on the end-of-life treatment, Figure \ref{fig:co2}. Under landfilling, increasing the number of uses reduces life-cycle emissions, as the production burden is distributed over a greater number of trips. Under incineration, the relationship reverses because the accounting convention assigns negative emissions to burning wood as a substitute for fossil fuels. Thus, the same change in $N$ can move the model output in opposite directions depending on the end-of-life state. 

\begin{center}
    \includegraphics[width=0.6\textwidth]{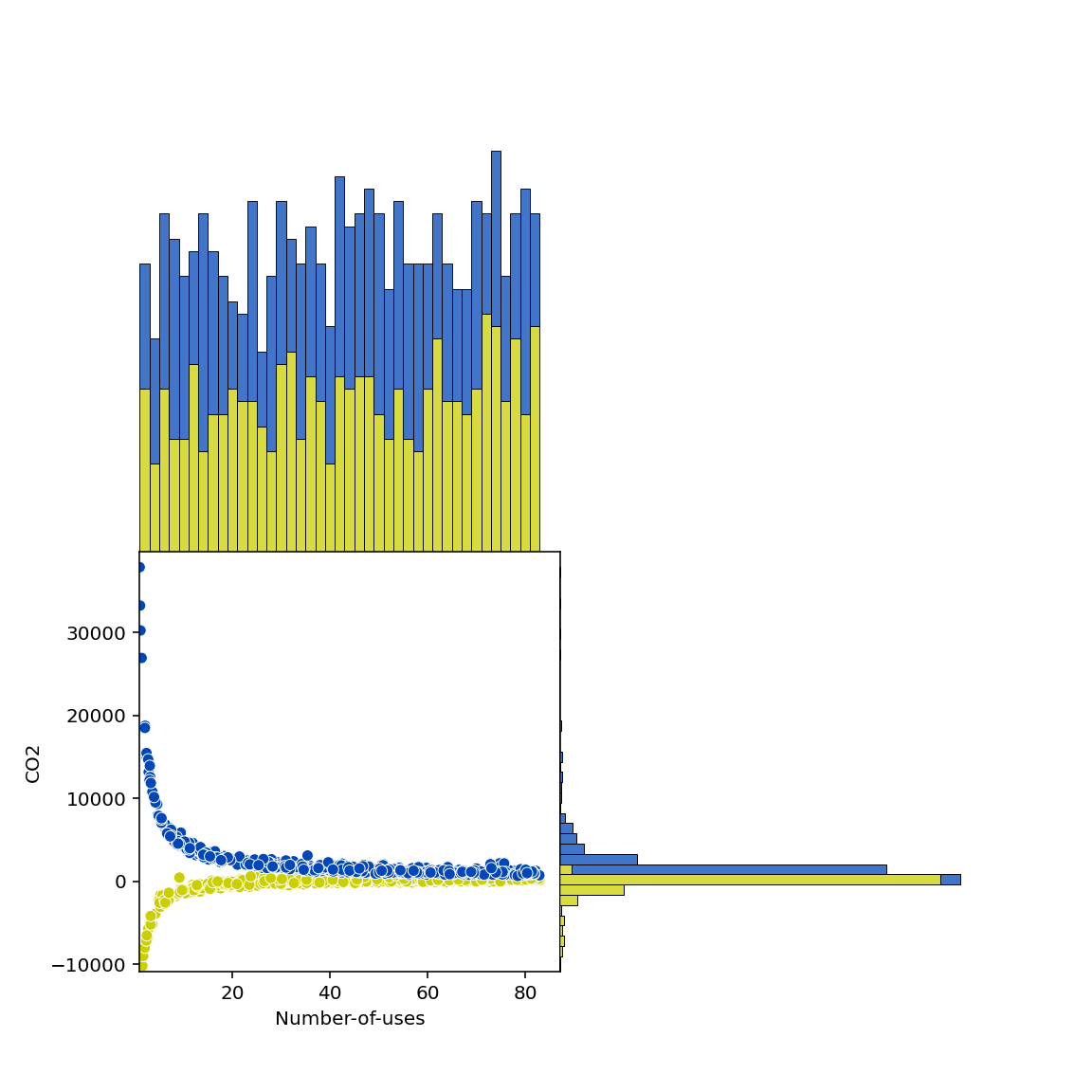}

    \captionof{figure}{
    Illustration of the reversal in the direction of the Number-of-uses effect on marginal CO$_2$ emissions depending on the End-of-life scenario: incineration (yellow) or landfilling (blue). 
    }
    \label{fig:co2}
\end{center}

Here, the model is re-simulated with $10^5$ Monte Carlo runs to obtain more stable estimates of both global sensitivity and heterogeneity, both are reported in Figure~\ref{fig:c1}.

\begin{center}
    \includegraphics[width=0.95\textwidth]{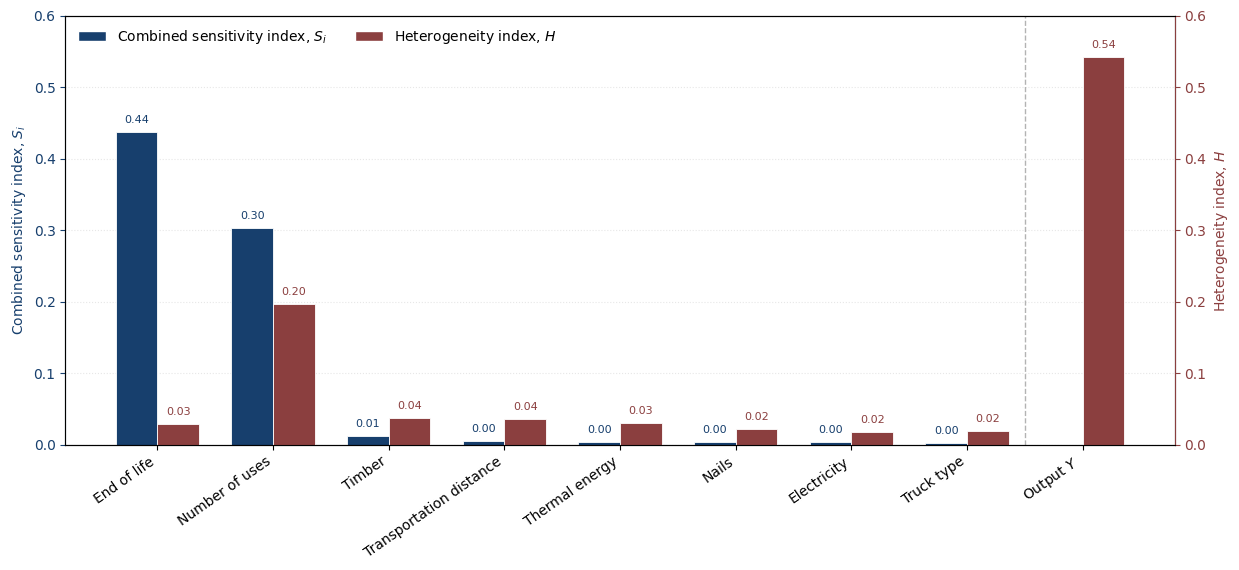}

    \captionof{figure}{
    Combined sensitivity and heterogeneity indices for the wooden-pallet carbon-footprint model.
    }
    \label{fig:c1}
\end{center}

Figure~\ref{fig:c1} reveals a markedly different picture when global importance and heterogeneity are considered jointly. End of life is the globally dominant variable ($S_i=0.44$), but its input-conditioned heterogeneity is very small ($H_{X_i}=0.03$). Its regional profile in Figure~\ref{fig:c2} (right) confirms the lack of variability in sensitivity between the two values. Hence, a reversal in the \emph{direction} of the effect, as expected, escapes the heterogeneity index, since variance-based sensitivity indices measure the magnitude of an effect rather than its sign. The heterogeneity index should therefore be interpreted as a screening measure for changes in sensitivity structure, rather than as a replacement for inspecting the underlying regional relationships.

\begin{center}
    \includegraphics[width=0.95\textwidth]{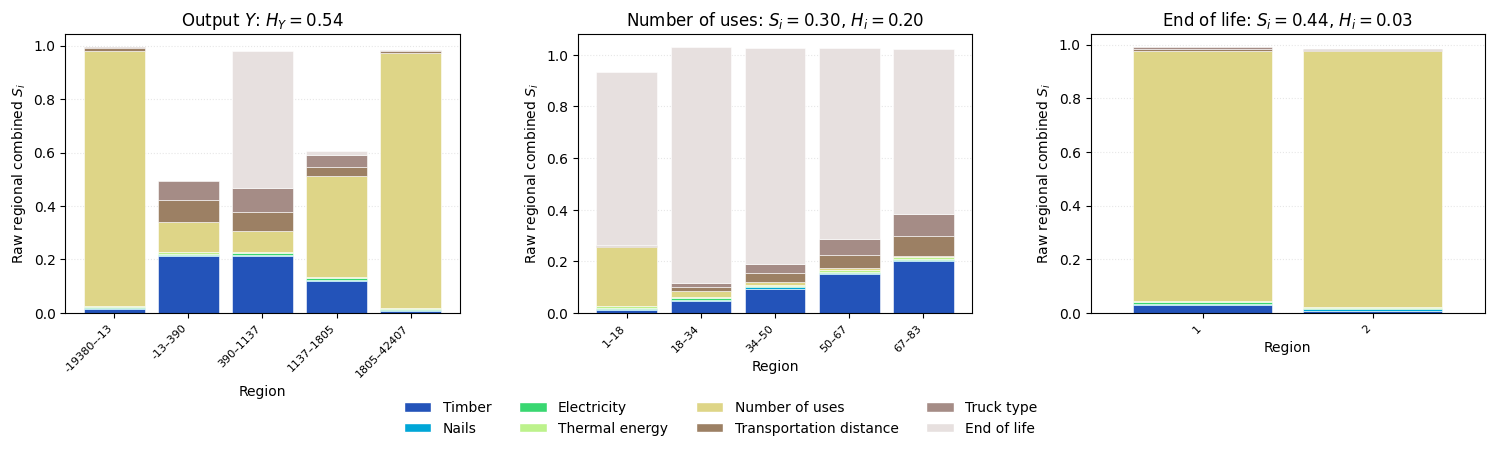}

    \captionof{figure}{
    Selected raw regional sensitivity profiles for the wooden-pallet carbon-footprint model. Profiles are shown for the output $Y$ and for inputs having either $H_{X_i}\geq0.10$ or $S_i\geq0.30$. The values of $H$ are calculated from normalized regional sensitivity profiles, while raw profiles are displayed here to retain information on regional explainability.
    }
    \label{fig:c2}
\end{center}

Nevertheless, the heterogeneity manifests elsewhere in the model, which was not detected in the previous analysis. The number of uses combines high global importance ($S_i=0.30$) with substantially greater heterogeneity ($H_{X_i}=0.20$). Its regional profile, Figure~\ref{fig:c2} (middle), shows the increasing effect of timber in higher regions, while the number of uses itself prevails only in the lowest region. 

Most strikingly, the output-conditioned heterogeneity is \(H_Y=0.54\), which indicates substantial changes in the composition of model sensitivity across the output distribution. Across the output space (Figure~\ref{fig:c2}, left), the number of uses dominates sensitivity in the extreme regions of the emissions distribution, while timber becomes influential in the intermediate regions despite its negligible global sensitivity index. End of life becomes the dominant contributor in the central region, precisely where the two end-of-life regimes meet and the effect of the number of uses reverses direction. Thus, although \(H_Y\) does not measure the sign of an effect directly, its large value captures the substantial redistribution of sensitivity associated with this reversal.

In practical terms, $H$ therefore provides a compact indication of when the global ranking is likely to be insufficient and when examining the regional sensitivity profiles is worthwhile.

\subsection{Flood model}
\label{sec:case_flood}

The second case revisits the canonical flood model examined extensively in the literature \citep{de2006maitrise,limbourg2010uncertainty,de2012modelling,Iooss2015Review,Baudin2016}, including a recent study on regional variance-based sensitivity analysis \citep{Helo2026Regional}. The model describes water depth $H_W$, overflow margin $S$, and total cost $C$ as functions of eight uncertain hydraulic and geometric inputs. In compact form,
\begin{equation}
    \alpha=\frac{Z_m-Z_v}{L},
    \qquad
    H_W=
    \left(
    \frac{Q}{K_sB\sqrt{\alpha}}
    \right)^{3/5},
    \qquad
    S=Z_v+H_W-Z_b-H_d,
\end{equation}
where flooding occurs when $S>0$. The cost $C$ combines dyke construction cost and flood-related cost, with the latter switching to a fixed flood cost when $S>0$. The full model formulation is given in \citet{Helo2026Regional}. The input assumptions used here are summarized in Table~\ref{tab:flood_inputs}.

\begin{center}
\captionof{table}{Uncertain inputs of the flood model.}
\label{tab:flood_inputs}
\begin{tabular}{llll}
\toprule
Symbol & Input & Probability distribution & Unit \\
\midrule
$Q$   & Flow rate & Gumbel$(\gamma=1013,\beta=558)$, $Q>0$ & m$^3$/s \\
$K_s$ & Strickler coefficient & $\mathcal{N}(30,7.5^2)$, $K_s>0$ & m$^{1/3}$/s \\
$Z_v$ & Downstream bed elevation & $\mathcal{U}(49,51)$ & m \\
$Z_m$ & Upstream bed elevation & $\mathcal{U}(54,56)$ & m \\
$B$   & River width & Triangular$(295,300,305)$ & m \\
$L$   & River reach length & Triangular$(4990,5000,5010)$ & m \\
$Z_b$ & Bank elevation & Triangular$(55,55.5,56)$ & m \\
$H_d$ & Dyke height & $\mathcal{U}(2,9)$ & m \\
\bottomrule
\end{tabular}
\end{center}

\paragraph{Standard heterogeneity analysis.}
The model was first evaluated using $N=100{,}000$ simulations and the standard heterogeneity analysis with $q=5$ equal-frequency regions. For each output, global combined sensitivity indices $S_i$, input-conditioned heterogeneity indices $H_{X_i}$, and output-conditioned heterogeneity $H_Y$ were calculated (Figure~\ref{fig:c3}).

\begin{center}
\includegraphics[width=0.99\textwidth]{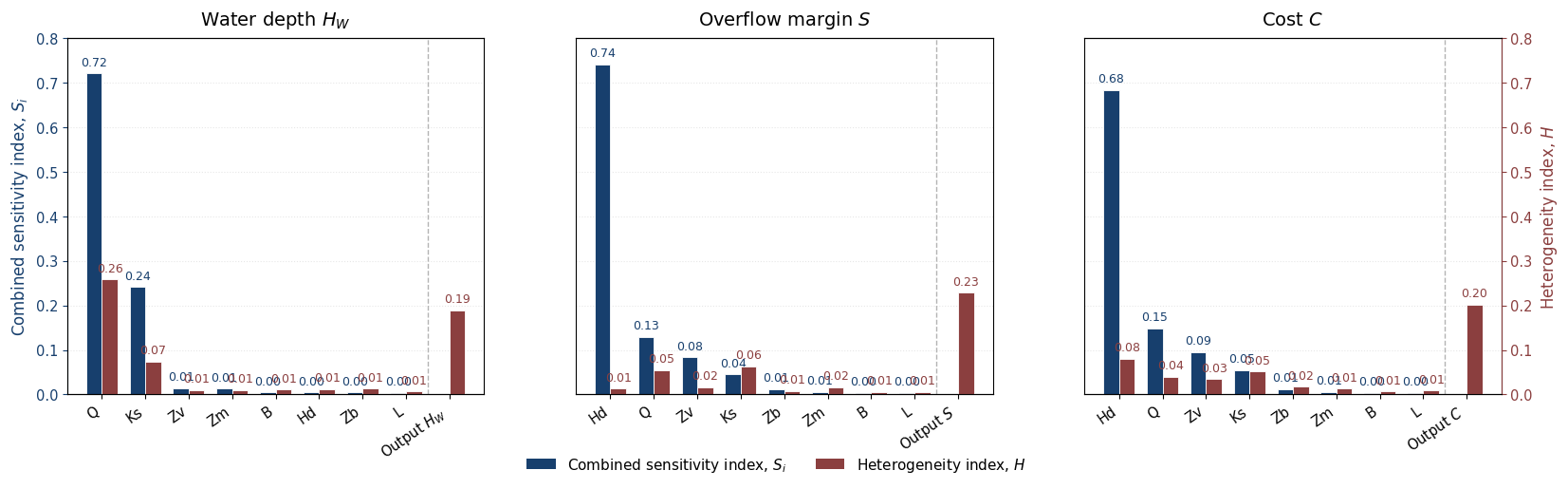}
\captionof{figure}{
Global combined sensitivity indices $S_i$ and heterogeneity indices for the flood model with (a) water depth $H_W$, (b) overflow margin $S$, and (c) cost $C$ as outputs. 
}
\label{fig:c3}
\end{center}

Overall, the standard analysis indicates only modest heterogeneity. Output-conditioned heterogeneity is $H_Y=0.19$ for water depth, $0.23$ for overflow, and $0.20$ for cost. Most input-conditioned indices are below 0.1, with the clearest exception being flow rate for water depth ($H_Q=0.26$).

These aggregate results are consistent with, but substantially more compact than, the earlier regional analysis that involved visual examination of regional sensitivity profiles \citep{Helo2026Regional}. That analysis showed that water depth is dominated by $Q$ at low values and by $K_s$ at high values; overflow changes from contributions by $Q$, $Z_v$, and $H_d$ in the lower tail to dominance by $H_d$ in intermediate regions and $K_s$ in the upper tail; and cost follows a broadly similar pattern, with $Q$ becoming particularly important in its upper region. The present $H_Y$ values quantify the overall magnitude of these previously observed redistributions. Similarly, the earlier input-conditioned analysis found the sensitivity profiles of $H_W$ and $S$ to be nearly invariant across $H_d$, but identified greater variation for $C$. The corresponding heterogeneity indices here, approximately $H_{H_d}=0.01$, $0.01$, and $0.08$, respectively, provide a concise quantitative confirmation of that result.

The modest values of $H_Y$, however, also expose a limitation of equal-frequency partitioning for rare-event models. Flood events occupy only a very small fraction of the simulation sample and are therefore diluted within a much larger upper quantile. A behaviorally important regime can consequently differ strongly from the remainder of the system while producing only moderate heterogeneity across equal-frequency regions.

\paragraph{Flood versus OK regimes.}
To isolate this behavior, an application-specific partition was defined as
\begin{equation}
Z_{\mathrm{flood}}=
\begin{cases}
\mathrm{OK}, & S\leq0,\\
\mathrm{Flood}, & S>0.
\end{cases}
\end{equation}
A larger sample of $3\times10^6$ simulations was generated to provide sufficient observations in the rare Flood regime, which accounted for roughly 900-1,000 observations or about 0.03\% of the sample. The same partition $Z_{\mathrm{flood}}$ was applied separately to all three outputs. The analysis was repeated 10 times to assess the stability of the resulting heterogeneity estimates under rare-event sampling. Table~\ref{tab:flood_h} reports the results of this experiment, and Figure \ref{fig:c4} shows the associated regional sensitivity profiles for one of the ten experiments.

\begin{center}
\captionof{table}{Heterogeneity between the OK and Flood regimes.}
\label{tab:flood_h}
\begin{tabular}{lcccc}
\toprule
Output & Mean $H_{\mathrm{flood}}$ & SD & Min--max $H_{\mathrm{flood}}$ & 
Min--max $\sum S_i$ in Flood \\
\midrule
Water depth $H_W$   & 0.641 & 0.038 & 0.608--0.740 & 0.306--0.968 \\
Overflow margin $S$ & 0.676 & 0.029 & 0.635--0.731 & 0.305--0.909 \\
Cost $C$             & 0.367 & 0.033 & 0.296--0.406 & 0.481--0.741 \\
\bottomrule
\end{tabular}
\end{center}

\begin{center}
\includegraphics[width=0.99\textwidth]{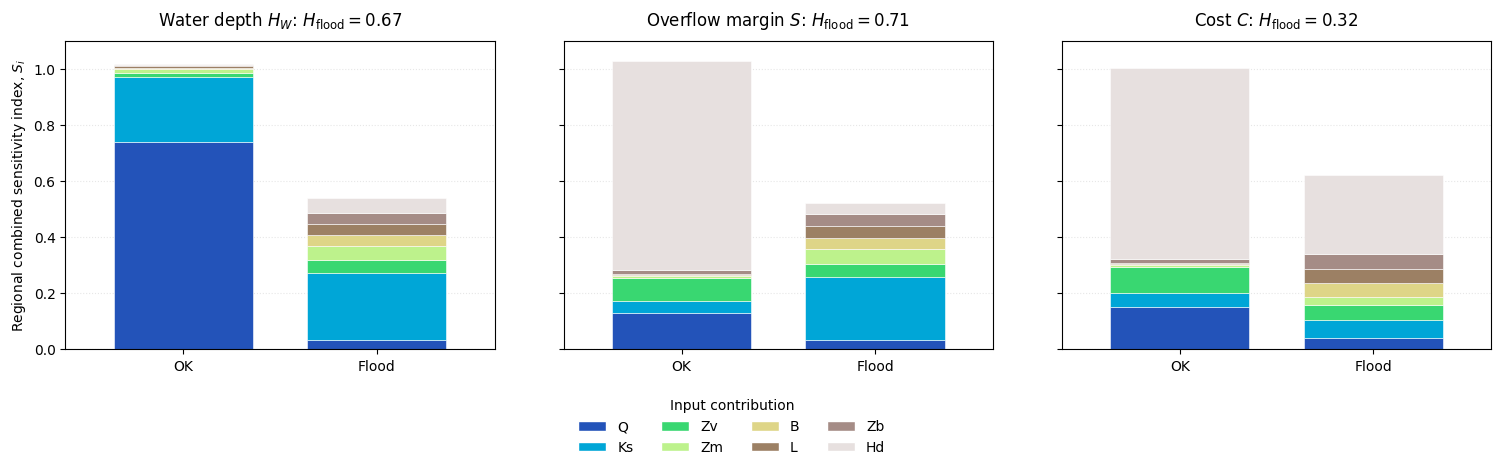}
\captionof{figure}{
Regional combined sensitivity profiles for the application-specific partition $Z_{\mathrm{flood}}=\{\mathrm{OK},\mathrm{Flood}\}$, shown for (a) water depth $H_W$, (b) overflow margin $S$, and (c) cost $C$. 
}
\label{fig:c4}
\end{center}

The resulting heterogeneity is considerably stronger than under equal-frequency partitioning. The Flood/OK distinction produces very high heterogeneity for water depth and overflow, with approximately two-thirds of the normalized sensitivity profile redistributed between the two regimes. Importantly, $H_{\mathrm{flood}}$ is high not only for overflow $S$, whose threshold defines the partition, but also for water depth $H_W$. The result therefore identifies a distinct behavioral region of the system in which the drivers of another continuous output change substantially.

The profiles in Figure~\ref{fig:c4} clarify the source of this large heterogeneity. In the OK regime, water-depth variance is overwhelmingly governed by $Q$, whereas within the Flood regime the influence shifts strongly toward $K_s$. A similar redistribution occurs for overflow: $H_d$, which is the dominant global source of variance, becomes much less important once flooding has already occurred, while $K_s$ becomes the largest contributor to variation within the Flood regime. Thus, a variable may be influential in determining whether the system enters a regime without remaining influential in explaining variation within that regime. The result extends the earlier regional study, which considered the Flood/No-flood partition only for $S$. It is also consistent with its quantile analysis, where $K_s$ became dominant in the highest overflow region \citep{Helo2026Regional}.

The raw sensitivity profiles also reveal an estimation issue in the rare Flood region. In the previous regional analysis, $N=10^5$ produced only 23 flood observations and a sensitivity-index sum above 8; increasing the sample to $10^6$ produced 316 flood observations and a sum close to one, which was interpreted as evidence of improved regional estimation \citep{Helo2026Regional}. The larger repeated experiments conducted here show that closeness of this sum to one is not, by itself, a robust convergence diagnostic for such an extreme region. Across repetitions, the raw sensitivity-index sum within Flood ranged from 0.306 to 0.968 for $H_W$ (mean $0.654$, SD $0.222$), from 0.305 to 0.909 for $S$ (mean $0.612$, SD $0.204$), and from 0.481 to 0.741 for $C$ (mean $0.552$, SD $0.082$), see the right column of Table \ref{tab:flood_h}. These deviations can reflect a combination of sampling variability in the rare and strongly skewed Flood region, dependencies induced by conditioning on the event, and variance associated with effects beyond the second order represented by the current combined indices.

In contrast, $H_{\mathrm{flood}}$ is substantially more stable across the same repetitions, particularly for $H_W$ and $S$ (Table~\ref{tab:flood_h}). This provides an empirical illustration of the role of normalization in the proposed index: $H_Z$ measures redistribution in the \emph{composition} of sensitivity rather than changes in the absolute regional sensitivity sum. The cost result should nevertheless be interpreted cautiously. Within the Flood regime, $C$ has very little remaining variance because the flood-related cost is fixed once $S>0$.

Overall, the case adds two insights to the previous regional analysis. First, the standard $H_Y$ and $H_{X_i}$ indices reduce a large collection of regional profiles to interpretable measures of the magnitude of heterogeneity while reproducing the qualitative conclusions of the earlier study. Second, the application-specific $H_{\mathrm{flood}}$ exposes strong behavioral heterogeneity that is largely diluted by equal-frequency regions. In particular, the transition from the overwhelmingly $Q$-driven OK regime to a $K_s$-dominated Flood regime is summarized by a single, high, and comparatively stable heterogeneity measure.

\subsection{Steel structure reliability model}
\label{sec:case_steel}

The third case revisits a structural reliability model for fatigue assessment of welded steel components that had previously been analyzed using global sensitivity analysis and Simulation Decomposition \citep{Kozlova2025Binning,ahola2024capturing}. The model is based on the 4R method, which combines material strength, welding residual stress, applied stress ratio, and weld quality to estimate the effective stress used in fatigue assessment \citep{ahola2021fatigue}. In simplified form, the output is the mean-stress-corrected reference effective notch stress range,
\begin{equation}
    \Delta\sigma_{k,\mathrm{ref}}
    =
    \frac{\Delta\sigma_k}
    {\sqrt{1-R_{\mathrm{local}}}},
\end{equation}
where the local stress state depends nonlinearly on the four model inputs. Thus, the model can be
written schematically as
\begin{equation}
    \Delta\sigma_{k,\mathrm{ref}}
    =
    f\!\left(
    \sigma_{\mathrm{res}},
    R,
    R_{p0.2},
    K_f
    \right).
\end{equation} 
which also illustrates the input-output set-up for sensitivity analysis: residual stress $\sigma_{\mathrm{res}}$, applied stress ratio $R$, steel grade $R_{p0.2}$, and the fatigue-effective stress factor $K_f$ are used as inputs. The original analysis used $10^4$ simulations and identified substantial nonlinear and heterogeneous relationships among the three most influential inputs \citep{Kozlova2025Binning,ahola2024capturing}.

Figure~\ref{fig:c5} shows that global importance and heterogeneity convey distinctly different information. Residual stress is both the globally dominant input ($S_i=0.52$) and strongly heterogeneous ($H_{\sigma_{\mathrm{res}}}=0.41$), indicating that its importance is accompanied by substantial changes in the surrounding sensitivity structure across its range. Stress ratio shows a similar, although somewhat weaker, combination of global importance and heterogeneity ($S_i=0.35$, $H_R=0.31$).

\begin{center}
    \includegraphics[width=0.95\textwidth]{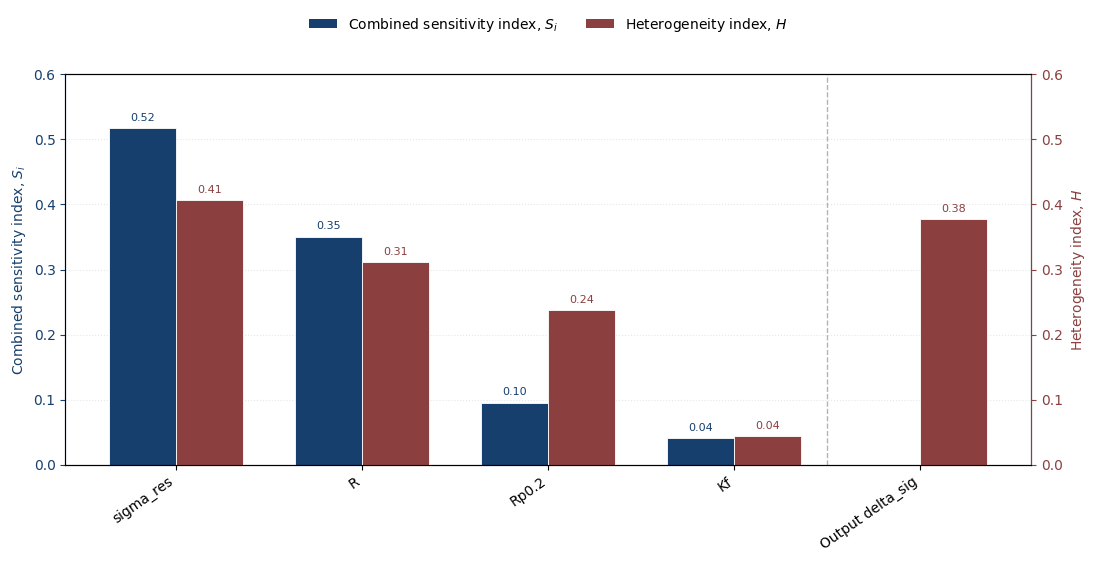}
    \captionof{figure}{
    Global combined sensitivity indices $S_i$ and heterogeneity indices for the steel structure reliability model. Input-conditioned heterogeneity $H_{X_i}$ is shown alongside the corresponding global sensitivity index, while $H_Y$ summarizes changes in sensitivity structure across the output distribution.
    }
    \label{fig:c5}
\end{center}

A different pattern is observed for steel grade. Its global sensitivity is relatively modest ($S_i=0.10$), yet its heterogeneity is considerably larger ($H_{R_{p0.2}}=0.24$). Thus, a ranking based only on global sensitivity would suggest a secondary role, whereas the heterogeneity index indicates that different values of steel grade are associated with substantially different sensitivity structures. This result is consistent with the earlier SimDec analysis, which showed that the effects of stress ratio and residual stress depend strongly on material-strength conditions \citep{Kozlova2025Binning,ahola2024capturing}. The output-conditioned heterogeneity is also substantial, $H_Y=0.38$, motivating inspection of the regional sensitivity profiles. 

Figure~\ref{fig:c6} reveals a third type of information that is not apparent from either global $S_i$ or input-conditioned $H_{X_i}$. The fatigue-effective stress factor $K_f$ has both negligible global sensitivity ($S_i=0.04$) and low input-conditioned heterogeneity ($H_{K_f}=0.04$). Nevertheless, its contribution changes markedly across the output distribution and it becomes the largest sensitivity contributor in the highest output region. Hence, an input that appears unimportant both globally and across its own range can still become locally important in a particular region of the model response.

\begin{center}
    \includegraphics[width=0.95\textwidth]{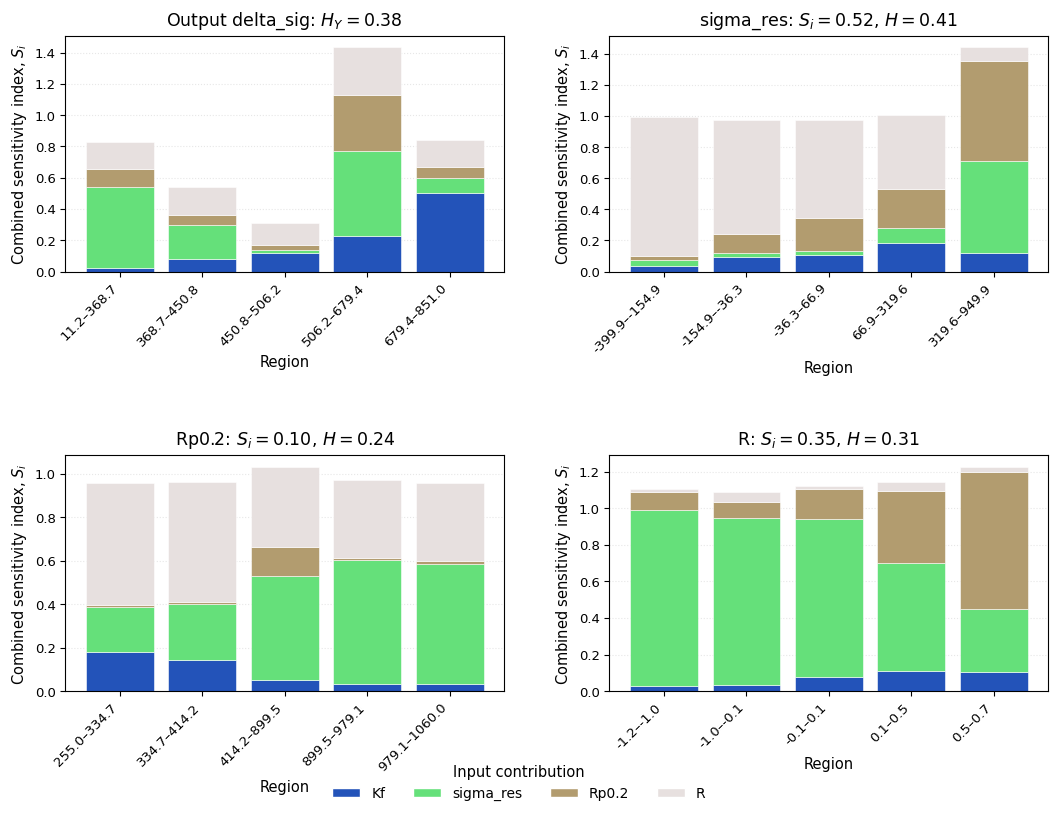}
    \captionof{figure}{
    Selected regional sensitivity profiles for the steel structure reliability model. Raw regional combined sensitivity indices are shown for the output and for inputs with $H_{X_i}\geq0.10$ or $S_i\geq0.30$. Heterogeneity indices are calculated from the corresponding normalized regional sensitivity profiles.
    }
    \label{fig:c6}
\end{center}

The case therefore illustrates three complementary situations: (i) residual stress is both globally important and heterogeneous; (ii) steel grade has modest global importance but comparatively high heterogeneity; and (iii) $K_f$ has low global and input-conditioned importance but becomes influential in an extreme output region. Together, these patterns demonstrate why $S_i$, $H_{X_i}$, and $H_Y$ each provide complementary views of model behavior.

\section{Conclusions}

This paper introduced the heterogeneity index, $H$, to quantify changes in the sensitivity profile across regions of a model. The index is based on the mean pairwise total-variation distance between normalized regional sensitivity profiles and is bounded between zero and one. Its general formulation, $H_P$, allows the regions to be defined by the output, yielding $H_Y$; by an input, yielding $H_{X_i}$; or by an application-specific partition variable, $Z$, representing meaningful behavioral or operational regimes.

Tests on controlled and benchmark models confirmed that the index behaves as intended: it approaches zero when regional sensitivity profiles remain stable, increases as deliberately introduced heterogeneity becomes stronger, and captures changes in sensitivity composition that are not apparent from global sensitivity indices alone. The experiments also showed that the number of regions and sample size affect estimation accuracy, but that the index becomes stable with sufficiently populated regions. Normalizing regional sensitivity profiles proved important for separating changes in the composition of sensitivity from changes in the absolute amount of regionally explained variance.

The three practical cases illustrate complementary ways in which this additional information matters. First, substantial output-conditioned heterogeneity can signal that a single global sensitivity profile conceals important changes in model behavior, including changes associated with reversals in the direction of effects. Second, an input with relatively modest global importance may nevertheless be associated with strong changes in the overall sensitivity structure. Third, application-specific partitions can reveal pronounced differences between behavioral regimes that are diluted by default equal-frequency regionalization, if the regime of interest is rare. Together, these cases demonstrate that global importance and heterogeneity answer different questions: $S_i$ indicates \emph{how much an input matters overall}, whereas $H$ indicates \emph{how much what matters changes}.

Our findings indicate that heterogeneity analysis should become a standard component of computational model assessment in conjunction with conventional sensitivity analysis. Its use is particularly important in high-stakes applications such as public policy, medicine, defence, aerospace, energy, and critical infrastructure, where decisions based on a single global sensitivity profile may overlook regime-dependent changes in what is driving the model outcomes. The proposed approach possesses an important limitation. The current implementation constructs regional sensitivity profiles from combined first- and second-order effects. Consequently, variance associated with third- and higher-order interactions is not represented. Future research can extend the framework toward higher-order effects while controlling the additional computational cost. Further work may also examine adaptive regionalization and uncertainty measures for $H$ in sparsely populated or rare-event regimes.

The heterogeneity-index function is implemented in the free and open-source \texttt{simdec} package \citep{simdec2026github} to facilitate straightforward application alongside existing sensitivity-analysis workflows and support its wider dissemination.

\section*{Software and data availability}

The heterogeneity index is implemented in the open-source Python package
\texttt{simdec}, version 1.5.2. The source code is publicly available at
\url{https://github.com/Simulation-Decomposition/simdec-python}, and the
package can be installed from PyPI using \texttt{pip install simdec}.

All computational experiments, case-study analyses, and figures reported in this paper can be reproduced using the Jupyter notebooks available at
\url{https://github.com/gnopik/Heterogeneity-index}.

\section*{Acknowledgements}

The authors gratefully acknowledge financial support from the Natural Sciences and Engineering Research Council of Canada (NSERC), Grant No.~OGP0155871.

\bibliographystyle{cas-model2-names}

\bibliography{REFERENCES}

\begin{thebibliography}{24}
\expandafter\ifx\csname natexlab\endcsname\relax\def\natexlab#1{#1}\fi
\providecommand{\url}[1]{\texttt{#1}}
\providecommand{\href}[2]{#2}
\providecommand{\path}[1]{#1}
\providecommand{\DOIprefix}{doi:}
\providecommand{\ArXivprefix}{arXiv:}
\providecommand{\URLprefix}{URL: }
\providecommand{\Pubmedprefix}{pmid:}
\providecommand{\doi}[1]{\href{http://dx.doi.org/#1}{\path{#1}}}
\providecommand{\Pubmed}[1]{\href{pmid:#1}{\path{#1}}}
\providecommand{\bibinfo}[2]{#2}
\ifx\xfnm\relax \def\xfnm[#1]{\unskip,\space#1}\fi
\bibitem[{Ahola et~al.(2024)Ahola, Kozlova and Yeomans}]{ahola2024capturing}
\bibinfo{author}{Ahola, A.}, \bibinfo{author}{Kozlova, M.},
  \bibinfo{author}{Yeomans, J.S.}, \bibinfo{year}{2024}.
\newblock \bibinfo{title}{Capturing multi-dimensional nonlinear behaviour of a
  steel structure reliability model--global sensitivity analysis}, in:
  \bibinfo{editor}{Kozlova, M.}, \bibinfo{editor}{Yeomans, J.S.} (Eds.),
  \bibinfo{booktitle}{Sensitivity Analysis for Business, Technology, and
  Policymaking Made Easy with Simulation Decomposition}.
  \bibinfo{publisher}{Routledge}. chapter~\bibinfo{chapter}{11}, pp.
  \bibinfo{pages}{257--282}.
\newblock \DOIprefix\doi{10.4324/9781003453789-15}.
\bibitem[{Ahola et~al.(2021)Ahola, Muikku, Braun and
  Bj{\"o}rk}]{ahola2021fatigue}
\bibinfo{author}{Ahola, A.}, \bibinfo{author}{Muikku, A.},
  \bibinfo{author}{Braun, M.}, \bibinfo{author}{Bj{\"o}rk, T.},
  \bibinfo{year}{2021}.
\newblock \bibinfo{title}{Fatigue strength assessment of ground fillet-welded
  joints using 4r method}.
\newblock \bibinfo{journal}{International journal of fatigue}
  \bibinfo{volume}{142}, \bibinfo{pages}{105916}.
\bibitem[{Azzini and Rosati(2022)}]{azzini2022function}
\bibinfo{author}{Azzini, I.}, \bibinfo{author}{Rosati, R.},
  \bibinfo{year}{2022}.
\newblock \bibinfo{title}{A function dataset for benchmarking in sensitivity
  analysis}.
\newblock \bibinfo{journal}{Data in Brief} \bibinfo{volume}{42},
  \bibinfo{pages}{108071}.
\newblock \DOIprefix\doi{10.1016/j.dib.2022.108071}.
\bibitem[{Baudin et~al.(2016)Baudin, Dutfoy, Iooss and Popelin}]{Baudin2016}
\bibinfo{author}{Baudin, M.}, \bibinfo{author}{Dutfoy, A.},
  \bibinfo{author}{Iooss, B.}, \bibinfo{author}{Popelin, A.L.},
  \bibinfo{year}{2016}.
\newblock \bibinfo{title}{OpenTURNS: An Industrial Software for Uncertainty
  Quantification in Simulation}. \bibinfo{publisher}{Springer International
  Publishing}, \bibinfo{address}{Cham}.
\newblock pp. \bibinfo{pages}{1--38}.
\newblock \URLprefix \url{https://doi.org/10.1007/978-3-319-11259-6_64-1},
  \DOIprefix\doi{10.1007/978-3-319-11259-6_64-1}.
\bibitem[{Beven and Freer(2001)}]{beven2001equifinality}
\bibinfo{author}{Beven, K.}, \bibinfo{author}{Freer, J.}, \bibinfo{year}{2001}.
\newblock \bibinfo{title}{Equifinality, data assimilation, and uncertainty
  estimation in mechanistic modelling of complex environmental systems using
  the glue methodology}.
\newblock \bibinfo{journal}{Journal of hydrology} \bibinfo{volume}{249},
  \bibinfo{pages}{11--29}.
\bibitem[{De~Rocquigny(2012)}]{de2012modelling}
\bibinfo{author}{De~Rocquigny, E.}, \bibinfo{year}{2012}.
\newblock \bibinfo{title}{Modelling under risk and uncertainty: an introduction
  to statistical, phenomenological and computational methods}.
\newblock \bibinfo{publisher}{John Wiley \& Sons}.
\bibitem[{Deviatkin et~al.(2021)Deviatkin, Kozlova and
  Yeomans}]{deviatkin2021simulation}
\bibinfo{author}{Deviatkin, I.}, \bibinfo{author}{Kozlova, M.},
  \bibinfo{author}{Yeomans, J.S.}, \bibinfo{year}{2021}.
\newblock \bibinfo{title}{Simulation decomposition for environmental
  sustainability: Enhanced decision-making in carbon footprint analysis}.
\newblock \bibinfo{journal}{Socio-Economic Planning Sciences}
  \bibinfo{volume}{75}, \bibinfo{pages}{100837}.
\bibitem[{Helo et~al.(2026)Helo, Kozlova, Roy and Yeomans}]{Helo2026Regional}
\bibinfo{author}{Helo, J.}, \bibinfo{author}{Kozlova, M.},
  \bibinfo{author}{Roy, P.}, \bibinfo{author}{Yeomans, J.S.},
  \bibinfo{year}{2026}.
\newblock \bibinfo{title}{Regional variance-based sensitivity analysis and a
  study of regional equifinality}.
\newblock \bibinfo{journal}{Risk Analysis} \bibinfo{volume}{46},
  \bibinfo{pages}{e70346}.
\bibitem[{Herman et~al.(2013)Herman, Reed and Wagener}]{herman2013time}
\bibinfo{author}{Herman, J.}, \bibinfo{author}{Reed, P.},
  \bibinfo{author}{Wagener, T.}, \bibinfo{year}{2013}.
\newblock \bibinfo{title}{Time-varying sensitivity analysis clarifies the
  effects of watershed model formulation on model behavior}.
\newblock \bibinfo{journal}{Water Resources Research} \bibinfo{volume}{49},
  \bibinfo{pages}{1400--1414}.
\bibitem[{Iooss and Lema{\^i}tre(2015)}]{Iooss2015Review}
\bibinfo{author}{Iooss, B.}, \bibinfo{author}{Lema{\^i}tre, P.},
  \bibinfo{year}{2015}.
\newblock \bibinfo{title}{A review on global sensitivity analysis methods}, in:
  \bibinfo{editor}{Dellino, G.}, \bibinfo{editor}{Meloni, C.} (Eds.),
  \bibinfo{booktitle}{Uncertainty Management in Simulation-Optimization of
  Complex Systems}. \bibinfo{publisher}{Springer}, \bibinfo{address}{Boston,
  MA}, pp. \bibinfo{pages}{101--122}.
\newblock \DOIprefix\doi{10.1007/978-1-4899-7547-8_5}.
\bibitem[{Kozlova et~al.(2025)Kozlova, Ahola, Roy and
  Yeomans}]{Kozlova2025Binning}
\bibinfo{author}{Kozlova, M.}, \bibinfo{author}{Ahola, A.},
  \bibinfo{author}{Roy, P.}, \bibinfo{author}{Yeomans, J.S.},
  \bibinfo{year}{2025}.
\newblock \bibinfo{title}{Simple binning algorithm and simdec visualization for
  comprehensive sensitivity analysis of complex computational models}.
\newblock \bibinfo{journal}{Journal of Environmental Informatics Letters}
  \bibinfo{volume}{13}, \bibinfo{pages}{38--56}.
\bibitem[{Kozlova et~al.(2024)Kozlova, Moss, Yeomans and
  Caers}]{Kozlova2024Heterogeneous}
\bibinfo{author}{Kozlova, M.}, \bibinfo{author}{Moss, R.J.},
  \bibinfo{author}{Yeomans, J.S.}, \bibinfo{author}{Caers, J.},
  \bibinfo{year}{2024}.
\newblock \bibinfo{title}{Uncovering heterogeneous effects in computational
  models for sustainable decision-making}.
\newblock \bibinfo{journal}{Environmental Modelling \& Software}
  \bibinfo{volume}{171}, \bibinfo{pages}{105898}.
\newblock \DOIprefix\doi{10.1016/j.envsoft.2023.105898}.
\bibitem[{Limbourg and De~Rocquigny(2010)}]{limbourg2010uncertainty}
\bibinfo{author}{Limbourg, P.}, \bibinfo{author}{De~Rocquigny, E.},
  \bibinfo{year}{2010}.
\newblock \bibinfo{title}{Uncertainty analysis using evidence
  theory--confronting level-1 and level-2 approaches with data availability and
  computational constraints}.
\newblock \bibinfo{journal}{Reliability Engineering \& System Safety}
  \bibinfo{volume}{95}, \bibinfo{pages}{550--564}.
\bibitem[{Marrel and Chabridon(2021)}]{marrel2021statistical}
\bibinfo{author}{Marrel, A.}, \bibinfo{author}{Chabridon, V.},
  \bibinfo{year}{2021}.
\newblock \bibinfo{title}{Statistical developments for target and conditional
  sensitivity analysis: Application on safety studies for nuclear reactor}.
\newblock \bibinfo{journal}{Reliability Engineering \& System Safety}
  \bibinfo{volume}{214}, \bibinfo{pages}{107711}.
\bibitem[{Massmann et~al.(2014)Massmann, Wagener and
  Holzmann}]{massmann2014new}
\bibinfo{author}{Massmann, C.}, \bibinfo{author}{Wagener, T.},
  \bibinfo{author}{Holzmann, H.}, \bibinfo{year}{2014}.
\newblock \bibinfo{title}{A new approach to visualizing time-varying
  sensitivity indices for environmental model diagnostics across evaluation
  time-scales}.
\newblock \bibinfo{journal}{Environmental Modelling \& Software}
  \bibinfo{volume}{51}, \bibinfo{pages}{190--194}.
\bibitem[{Pianosi et~al.(2016)Pianosi, Beven, Freer, Hall, Rougier, Stephenson
  and Wagener}]{pianosi2016sensitivity}
\bibinfo{author}{Pianosi, F.}, \bibinfo{author}{Beven, K.},
  \bibinfo{author}{Freer, J.}, \bibinfo{author}{Hall, J.W.},
  \bibinfo{author}{Rougier, J.}, \bibinfo{author}{Stephenson, D.B.},
  \bibinfo{author}{Wagener, T.}, \bibinfo{year}{2016}.
\newblock \bibinfo{title}{Sensitivity analysis of environmental models: A
  systematic review with practical workflow}.
\newblock \bibinfo{journal}{Environmental Modelling \& Software}
  \bibinfo{volume}{79}, \bibinfo{pages}{214--232}.
\bibitem[{de~Rocquigny(2006)}]{de2006maitrise}
\bibinfo{author}{de~Rocquigny, {\'E}.}, \bibinfo{year}{2006}.
\newblock \bibinfo{title}{La ma{\^\i}trise des incertitudes dans un contexte
  industriel. 1re partie: une approche m{\'e}thodologique globale bas{\'e}e sur
  des exemples}.
\newblock \bibinfo{journal}{Journal de la Soci{\'e}t{\'e} fran{\c{c}}aise de
  statistique} \bibinfo{volume}{147}, \bibinfo{pages}{33--71}.
\bibitem[{Roux et~al.(2021)Roux, Buis, Lafolie and Lamboni}]{roux2021cluster}
\bibinfo{author}{Roux, S.}, \bibinfo{author}{Buis, S.},
  \bibinfo{author}{Lafolie, F.}, \bibinfo{author}{Lamboni, M.},
  \bibinfo{year}{2021}.
\newblock \bibinfo{title}{Cluster-based gsa: Global sensitivity analysis of
  models with temporal or spatial outputs using clustering}.
\newblock \bibinfo{journal}{Environmental Modelling \& Software}
  \bibinfo{volume}{140}, \bibinfo{pages}{105046}.
\bibitem[{Roy and Kozlova(2024)}]{roy2024simulation}
\bibinfo{author}{Roy, P.T.}, \bibinfo{author}{Kozlova, M.},
  \bibinfo{year}{2024}.
\newblock \bibinfo{title}{Simulation decomposition in python}.
\newblock \bibinfo{journal}{Journal of Open Source Software}
  \bibinfo{volume}{9}, \bibinfo{pages}{6713}.
\bibitem[{{Simulation-Decomposition}(2026)}]{simdec2026github}
\bibinfo{author}{{Simulation-Decomposition}}, \bibinfo{year}{2026}.
\newblock \bibinfo{title}{Simulation-decomposition: Sensitivity analysis using
  simulation decomposition}.
\newblock
  \bibinfo{howpublished}{\url{https://github.com/Simulation-Decomposition}}.
\newblock \bibinfo{note}{Accessed: 2026-04-19}.
\bibitem[{Spear and Hornberger(1980)}]{spear1980eutrophication}
\bibinfo{author}{Spear, R.C.}, \bibinfo{author}{Hornberger, G.M.},
  \bibinfo{year}{1980}.
\newblock \bibinfo{title}{Eutrophication in peel inlet—ii. identification of
  critical uncertainties via generalized sensitivity analysis}.
\newblock \bibinfo{journal}{Water research} \bibinfo{volume}{14},
  \bibinfo{pages}{43--49}.
\bibitem[{Tarantola et~al.(2024)Tarantola, Ferretti, Lo~Piano, Kozlova, Lachi,
  Rosati, Puy, Roy, Vannucci, Kuc-Czarnecka and
  Saltelli}]{tarantola2024annotated}
\bibinfo{author}{Tarantola, S.}, \bibinfo{author}{Ferretti, F.},
  \bibinfo{author}{Lo~Piano, S.}, \bibinfo{author}{Kozlova, M.},
  \bibinfo{author}{Lachi, A.}, \bibinfo{author}{Rosati, R.},
  \bibinfo{author}{Puy, A.}, \bibinfo{author}{Roy, P.},
  \bibinfo{author}{Vannucci, G.}, \bibinfo{author}{Kuc-Czarnecka, M.},
  \bibinfo{author}{Saltelli, A.}, \bibinfo{year}{2024}.
\newblock \bibinfo{title}{An annotated timeline of sensitivity analysis}.
\newblock \bibinfo{journal}{Environmental Modelling \& Software}
  \bibinfo{volume}{174}, \bibinfo{pages}{105977}.
\bibitem[{Wei et~al.(2015)Wei, Lu and Song}]{wei2015regional}
\bibinfo{author}{Wei, P.}, \bibinfo{author}{Lu, Z.}, \bibinfo{author}{Song,
  J.}, \bibinfo{year}{2015}.
\newblock \bibinfo{title}{Regional and parametric sensitivity analysis of
  sobol' indices}.
\newblock \bibinfo{journal}{Reliability Engineering \& System Safety}
  \bibinfo{volume}{137}, \bibinfo{pages}{87--100}.
\bibitem[{Wei et~al.(2013)Wei, Lu, Wu and Zhou}]{wei2013moment}
\bibinfo{author}{Wei, P.}, \bibinfo{author}{Lu, Z.}, \bibinfo{author}{Wu, D.},
  \bibinfo{author}{Zhou, C.}, \bibinfo{year}{2013}.
\newblock \bibinfo{title}{Moment-independent regional sensitivity analysis:
  application to an environmental model}.
\newblock \bibinfo{journal}{Environmental modelling \& software}
  \bibinfo{volume}{47}, \bibinfo{pages}{55--63}.

\end{thebibliography}



\appendix

\section*{Appendix 1. Regional sensitivity profiles across different numbers of regions}
\label{app:regional_profiles}

\begin{center}
    \includegraphics[width=0.95\textwidth]{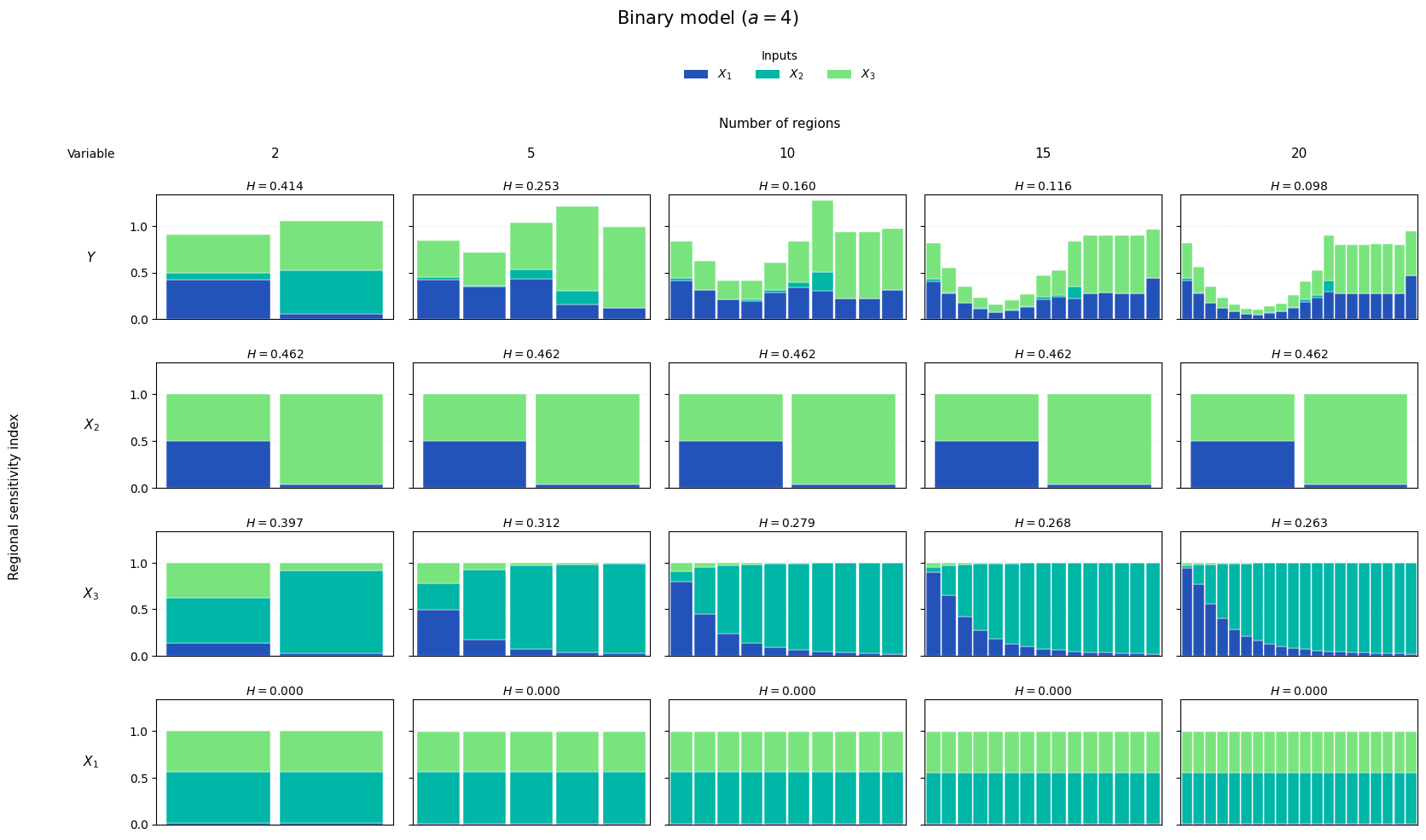}
    \captionof{figure}{Regional sensitivity profiles for the controlled binary model.}
    \label{fig:a1}
\end{center}

\begin{center}
    \includegraphics[width=0.95\textwidth]{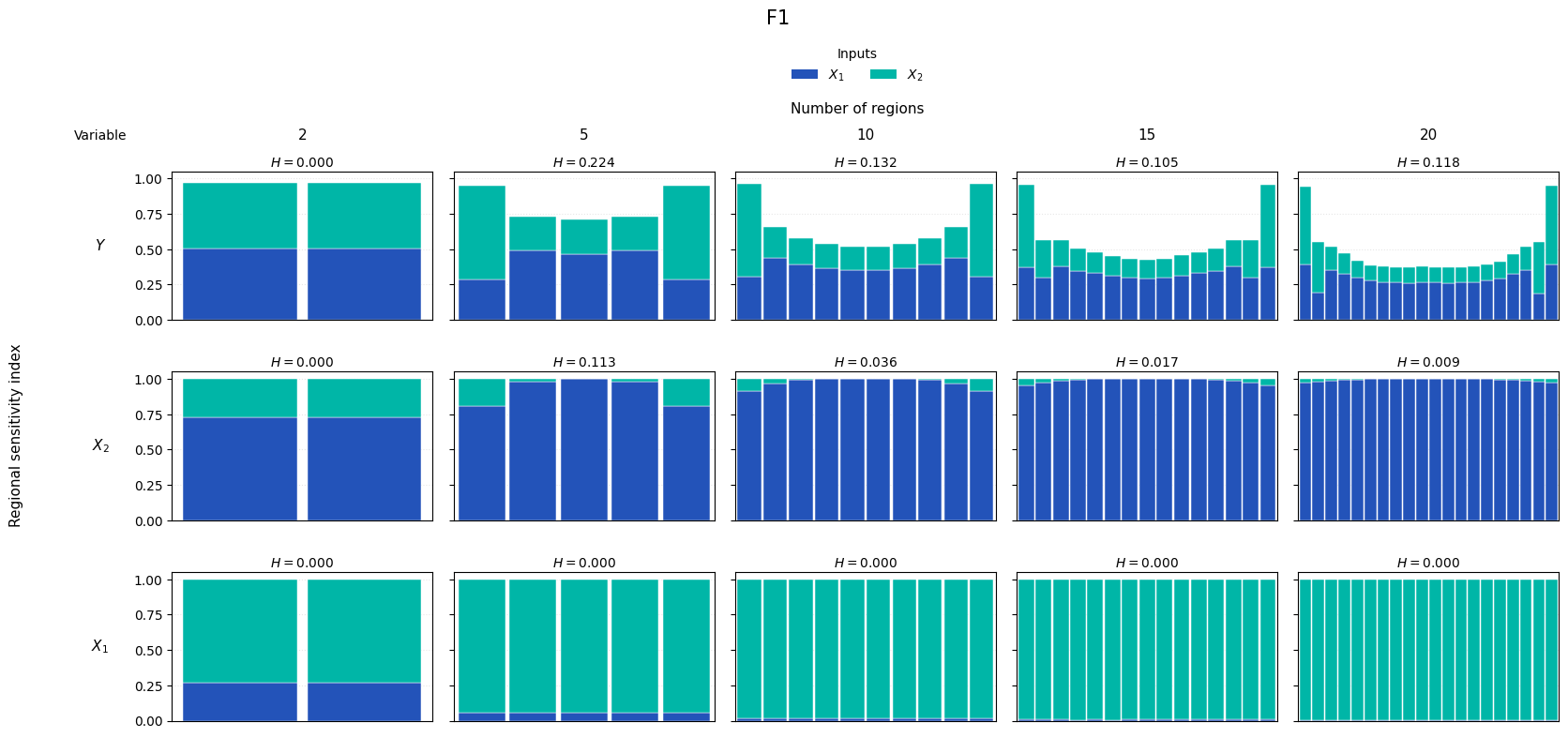}
    \captionof{figure}{Regional sensitivity profiles for F1 function.}
    \label{fig:a2}
\end{center}

\begin{center}
    \includegraphics[width=0.95\textwidth]{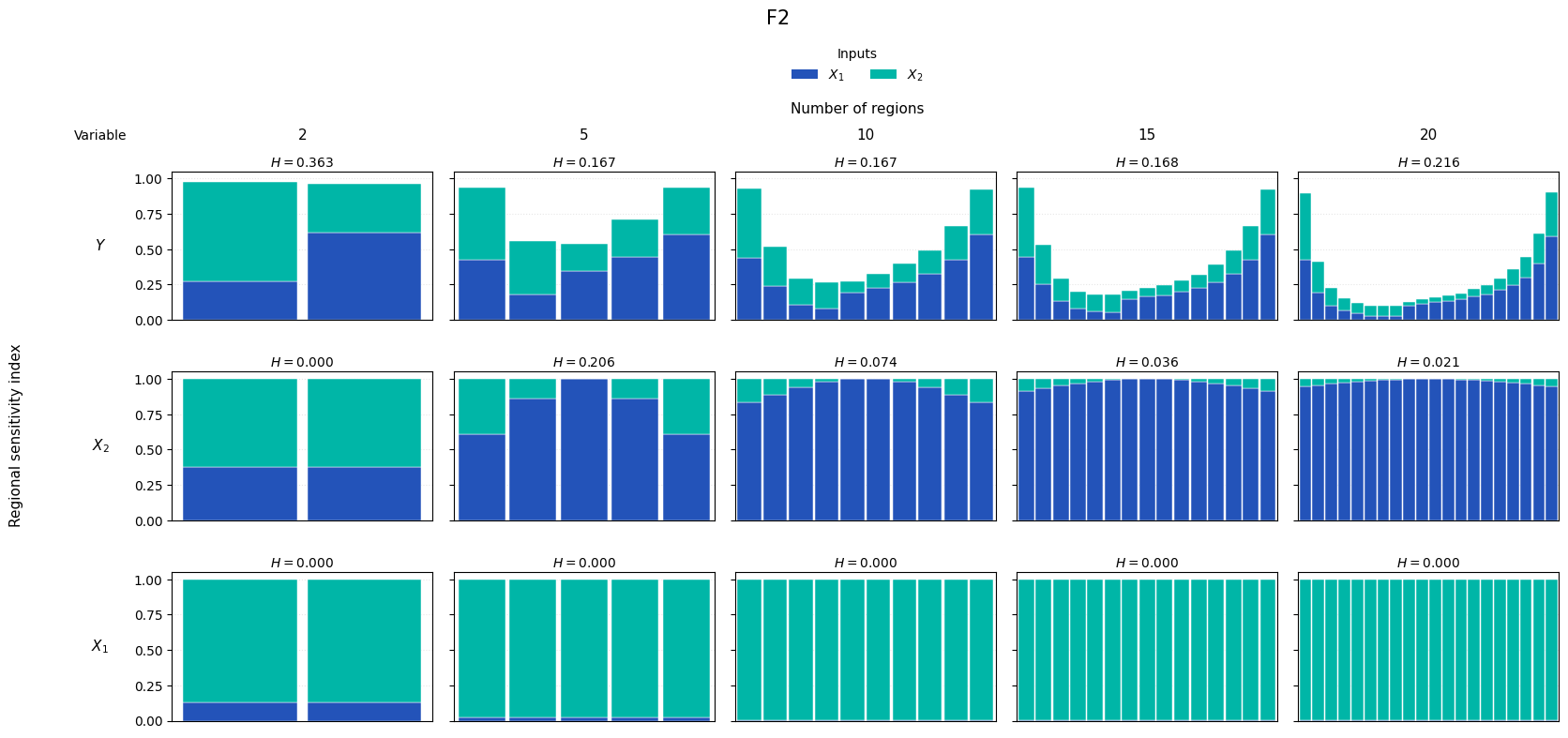}
    \captionof{figure}{Regional sensitivity profiles for F2 function.}
    \label{fig:a3}
\end{center}

\begin{center}
    \includegraphics[width=0.95\textwidth]{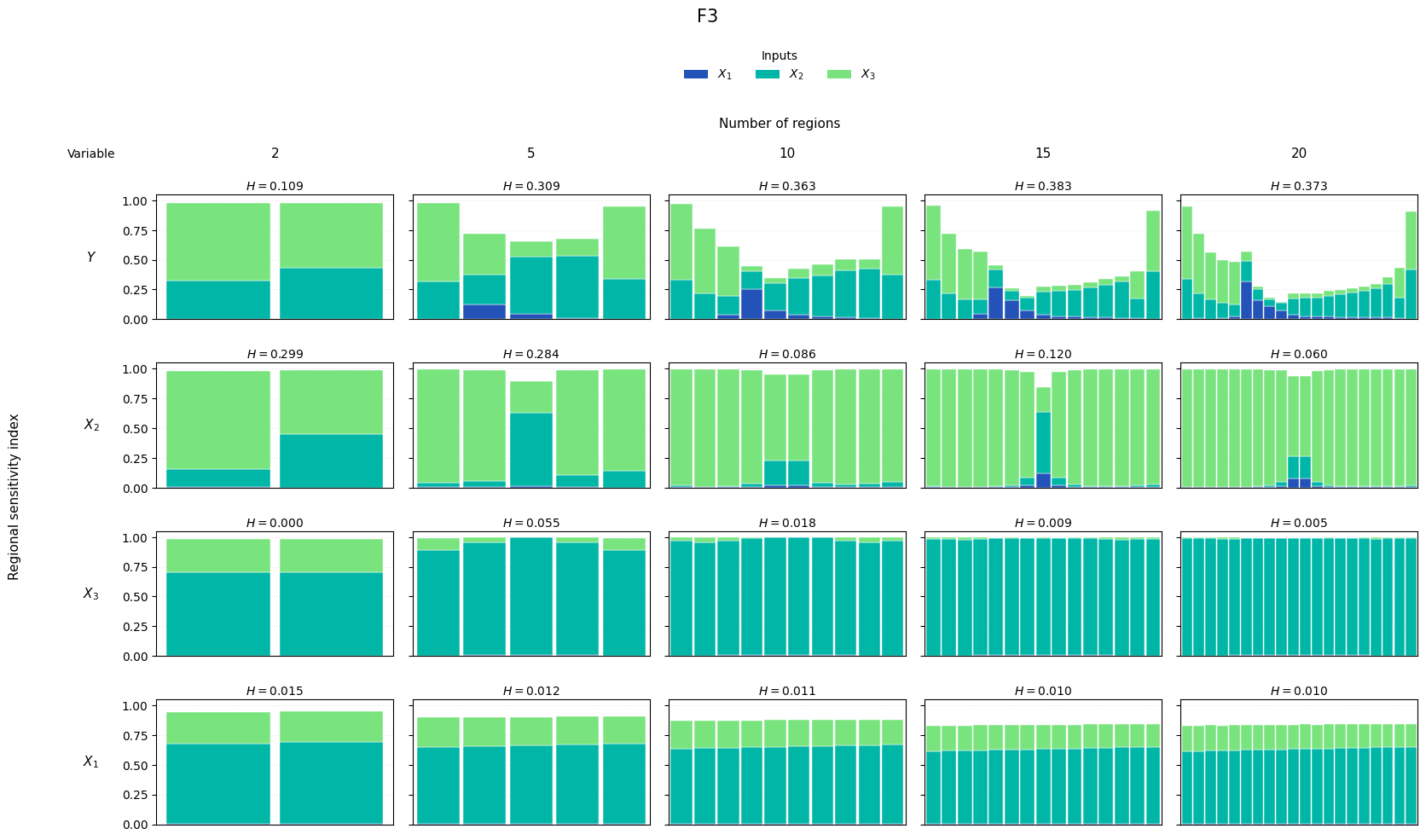}
    \captionof{figure}{Regional sensitivity profiles for F3 function.}
    \label{fig:a4}
\end{center}

\begin{center}
    \includegraphics[width=0.95\textwidth]{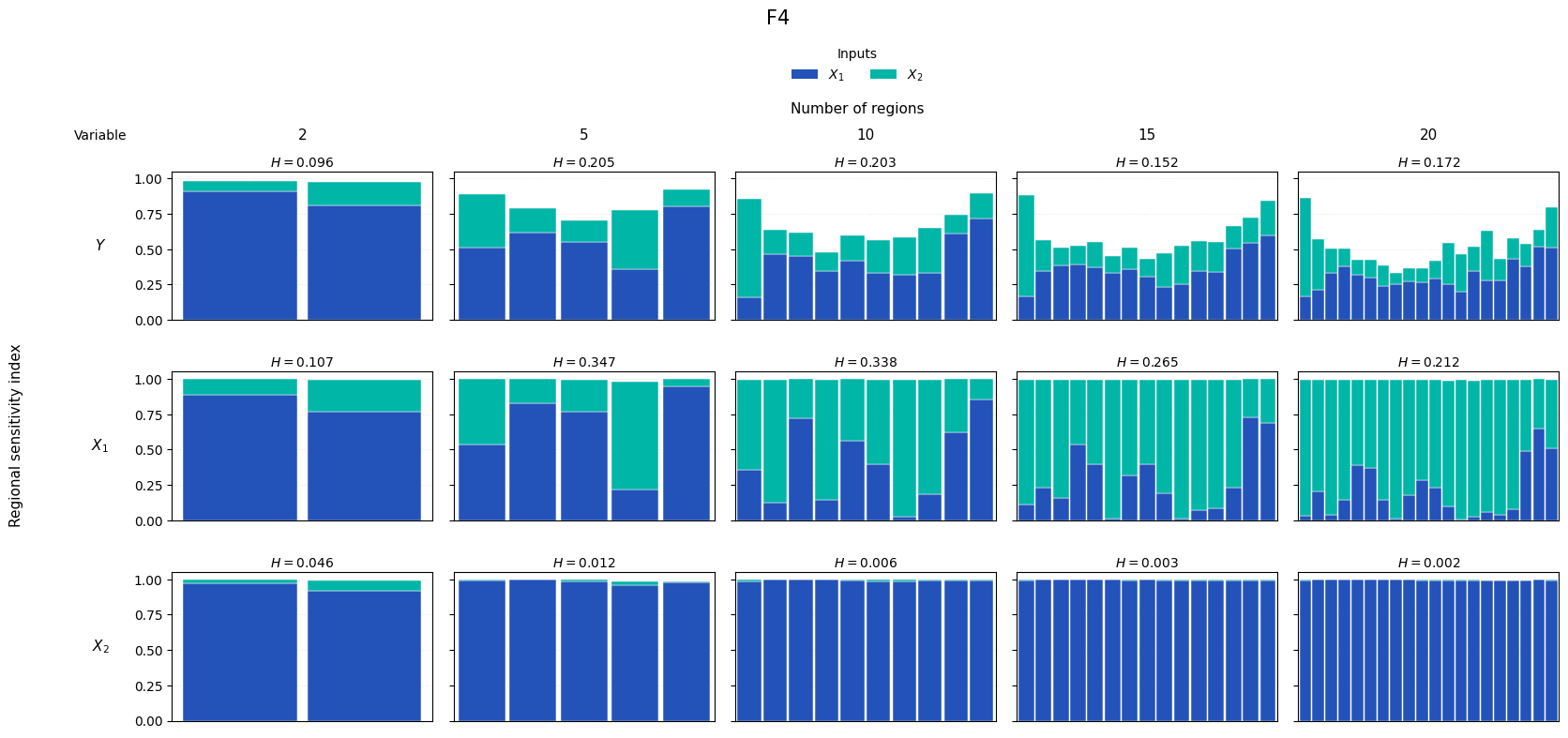}
    \captionof{figure}{Regional sensitivity profiles for F4 function.}
    \label{fig:a5}
\end{center}

\begin{center}
    \includegraphics[width=0.95\textwidth]{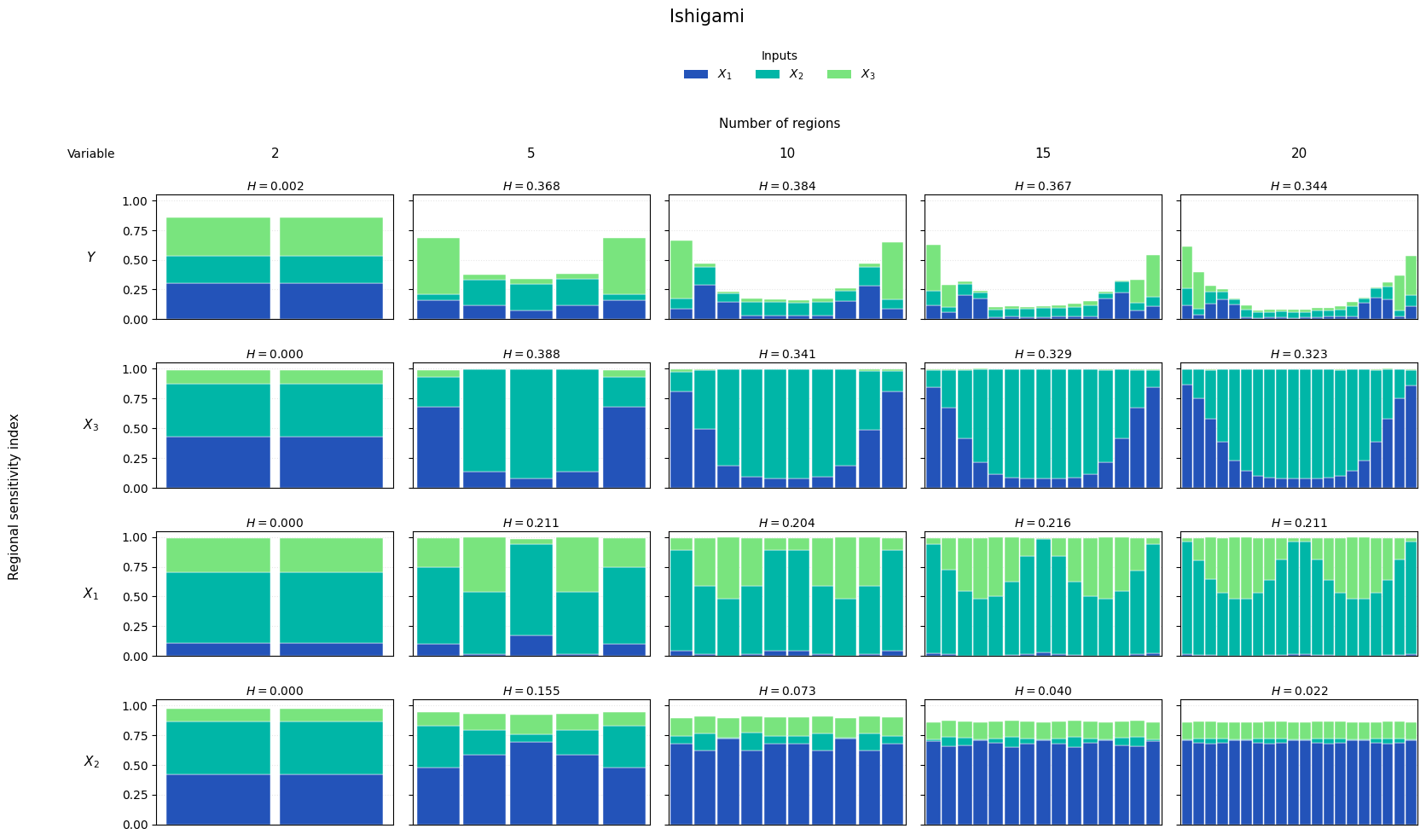}
    \captionof{figure}{Regional sensitivity profiles for the Ishigami function.}
    \label{fig:a6}
\end{center}

\begin{center}
    \includegraphics[width=0.95\textwidth]{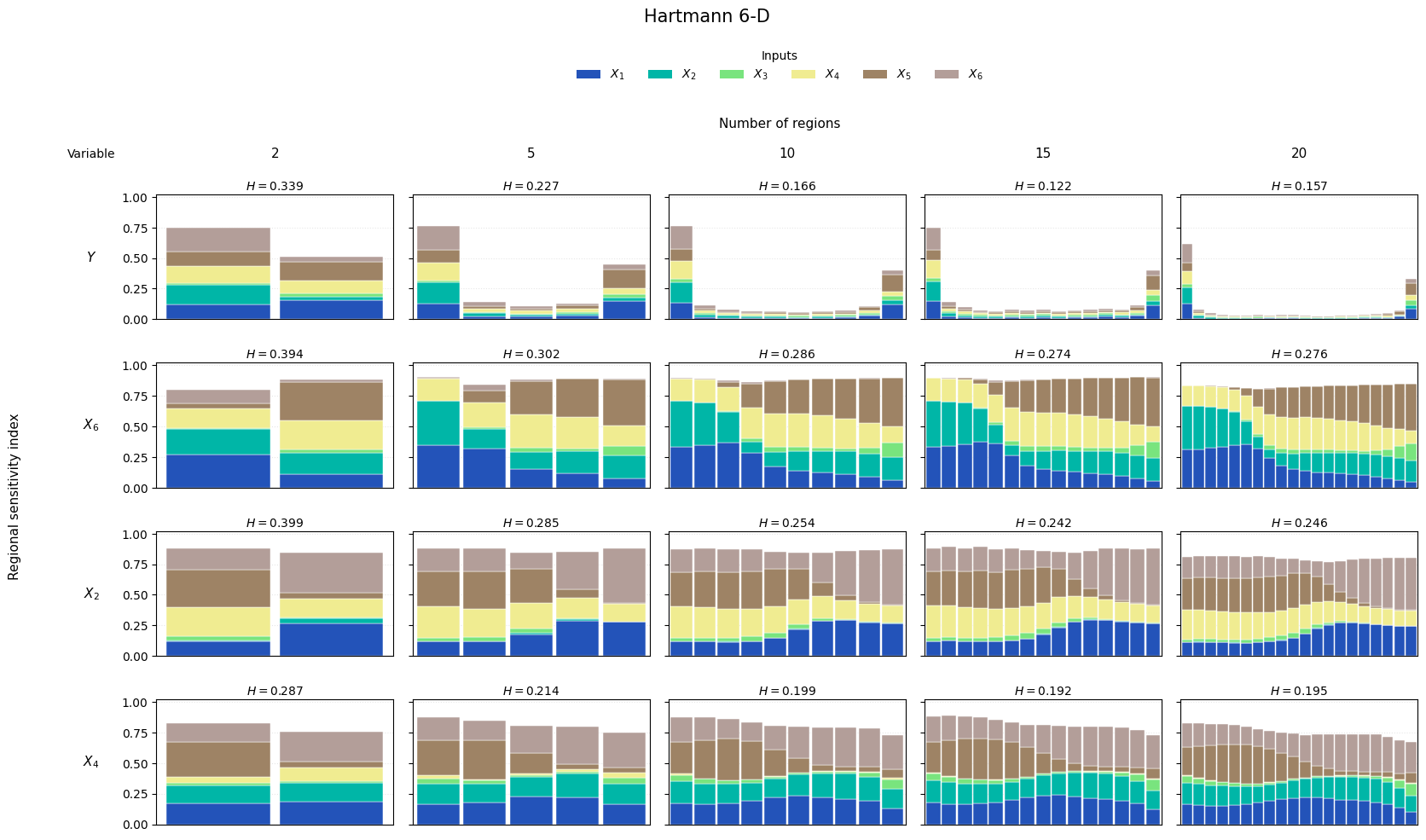}
    \captionof{figure}{Regional sensitivity profiles for the Hartmann 6-D function.}
    \label{fig:a7}
\end{center}

\begin{center}
    \includegraphics[width=0.95\textwidth]{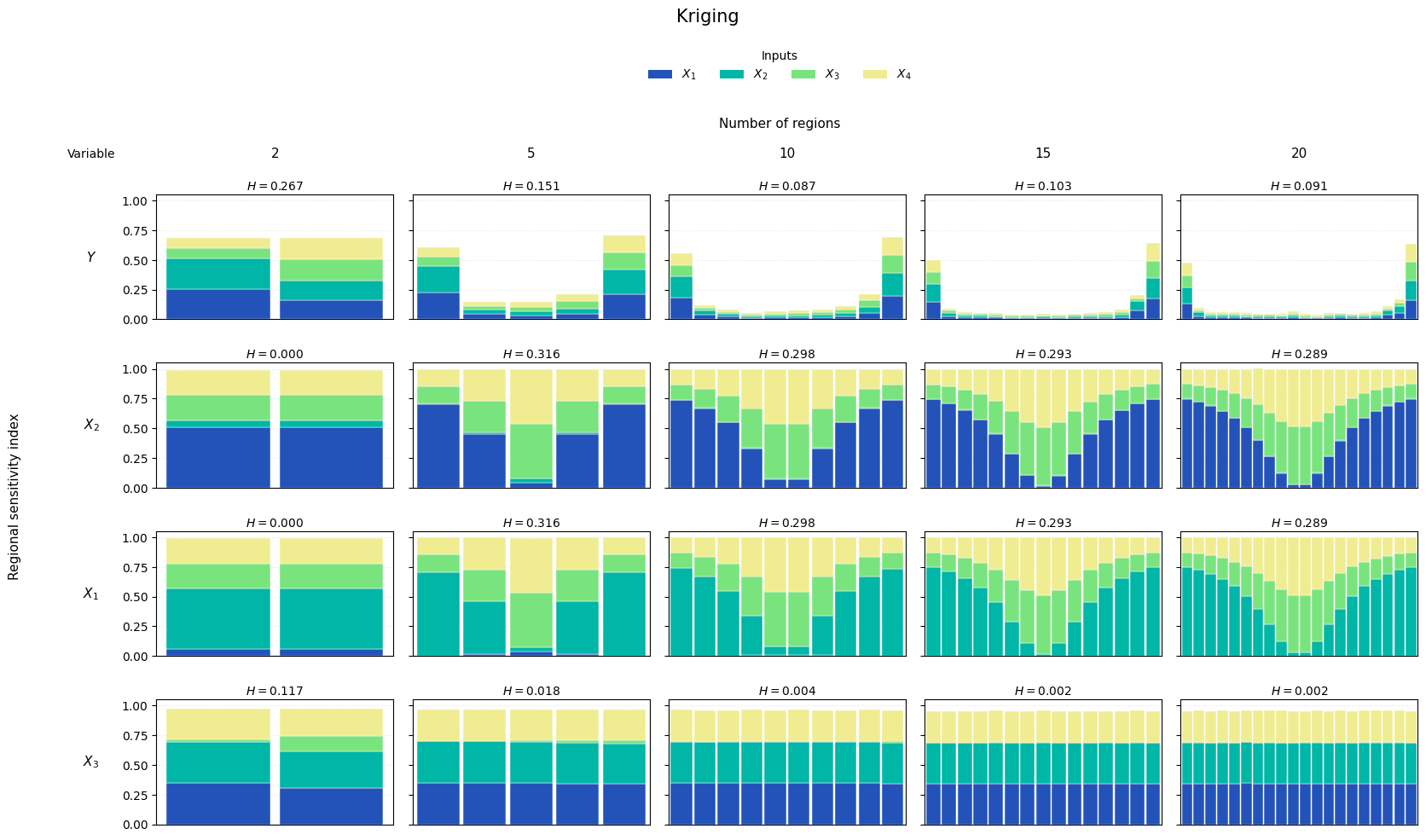}
    \captionof{figure}{Regional sensitivity profiles for the Kriging function.}
    \label{fig:a8}
\end{center}

\begin{center}
    \includegraphics[width=0.95\textwidth]{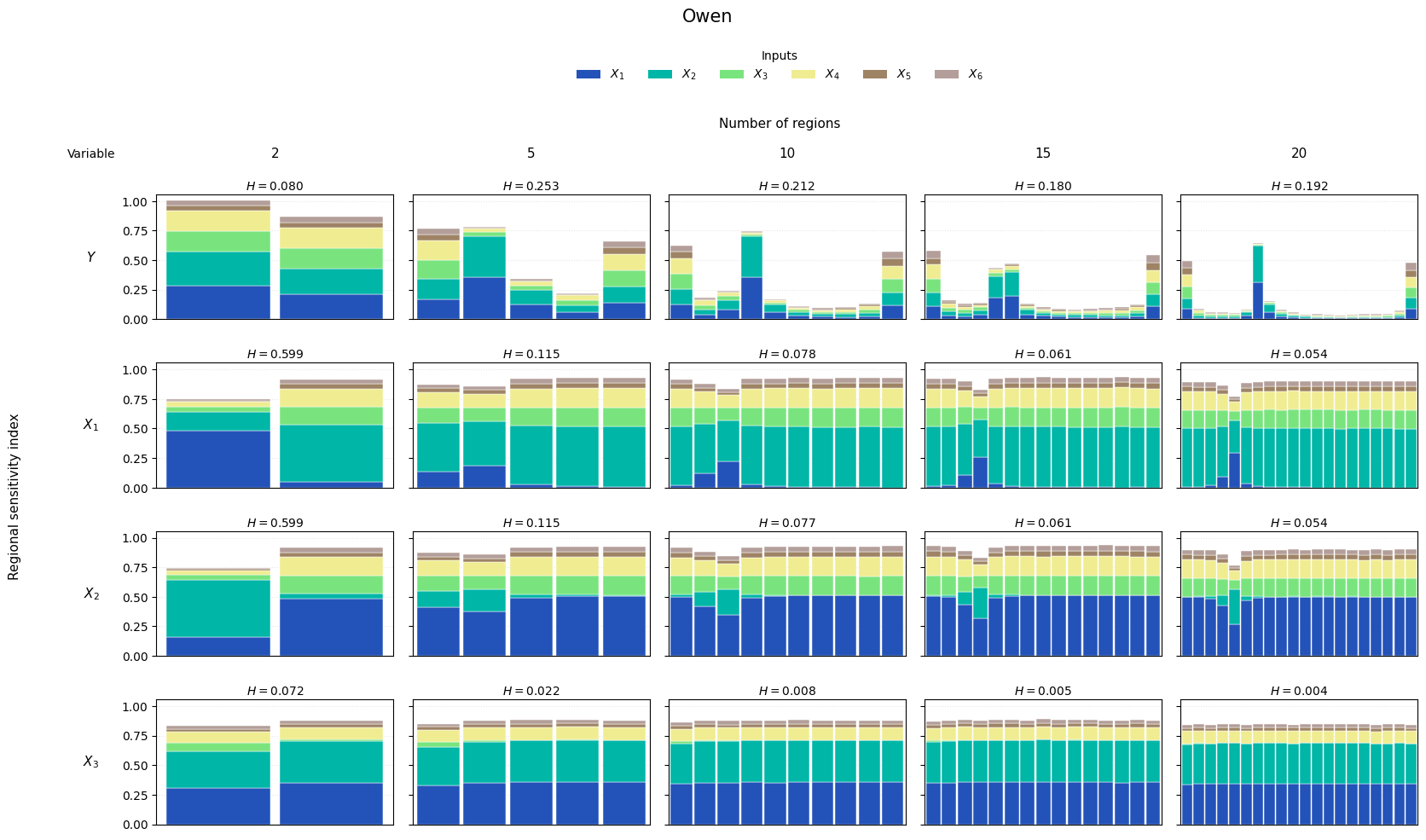}
    \captionof{figure}{Regional sensitivity profiles for the Owen function.}
    \label{fig:a9}
\end{center}

\clearpage

\section*{Appendix 2. Sample-size stability of the heterogeneity index}
\label{app:sample_size}

\begin{center}
    \includegraphics[width=0.95\textwidth]{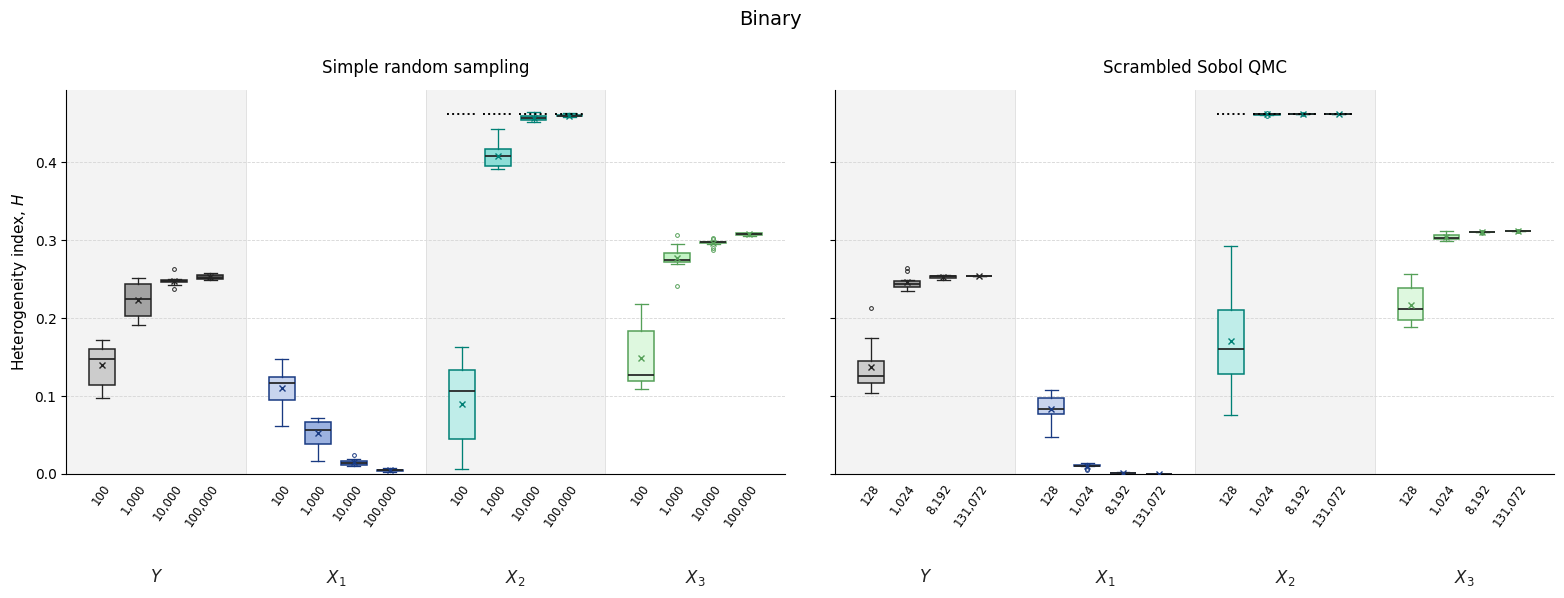}
    \captionof{figure}{Sample-size stability of the heterogeneity index $H$ for the controlled binary model under simple random sampling and scrambled Sobol' QMC sampling. The dotted reference line for $X_2=K$ indicates the analytical value $H_K=6/13\approx0.462$.}
    \label{fig:b1}
\end{center}

\begin{center}
    \includegraphics[width=0.95\textwidth]{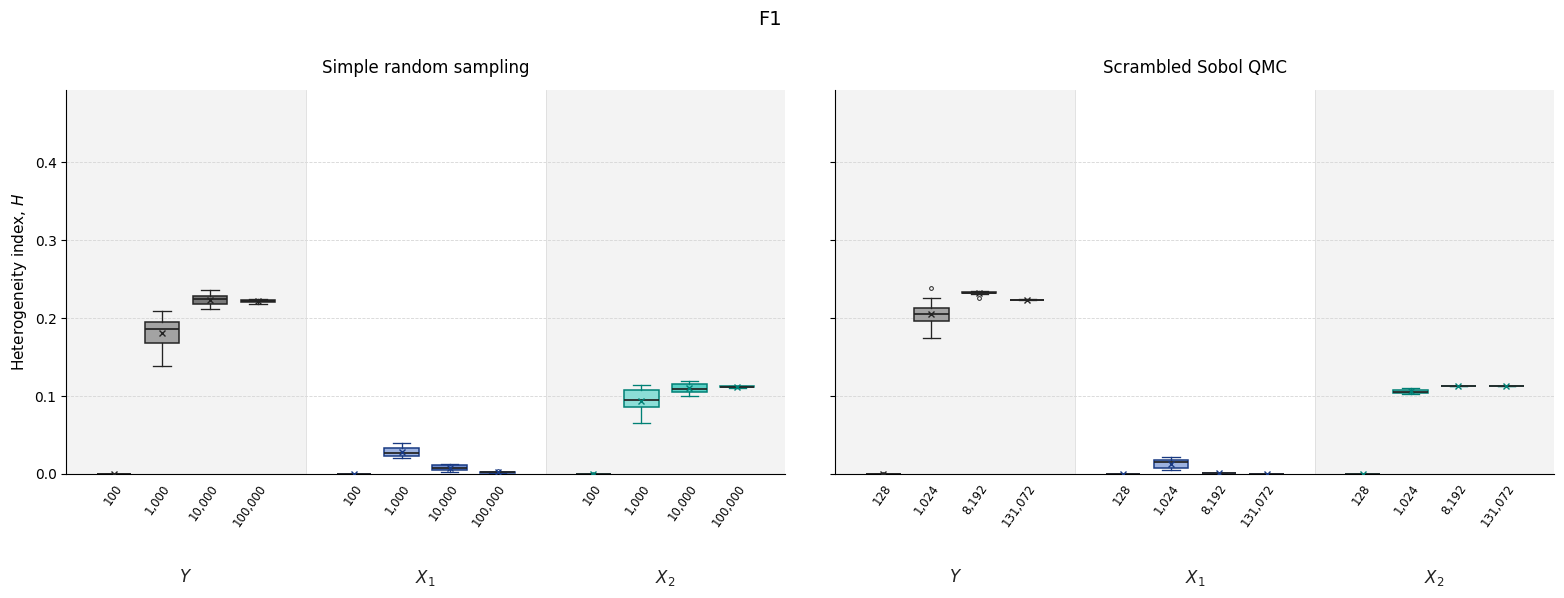}
    \captionof{figure}{Sample-size stability of the heterogeneity index $H$ for F1 function under simple random sampling and scrambled Sobol' QMC sampling.}
    \label{fig:b2}
\end{center}

\begin{center}
    \includegraphics[width=0.95\textwidth]{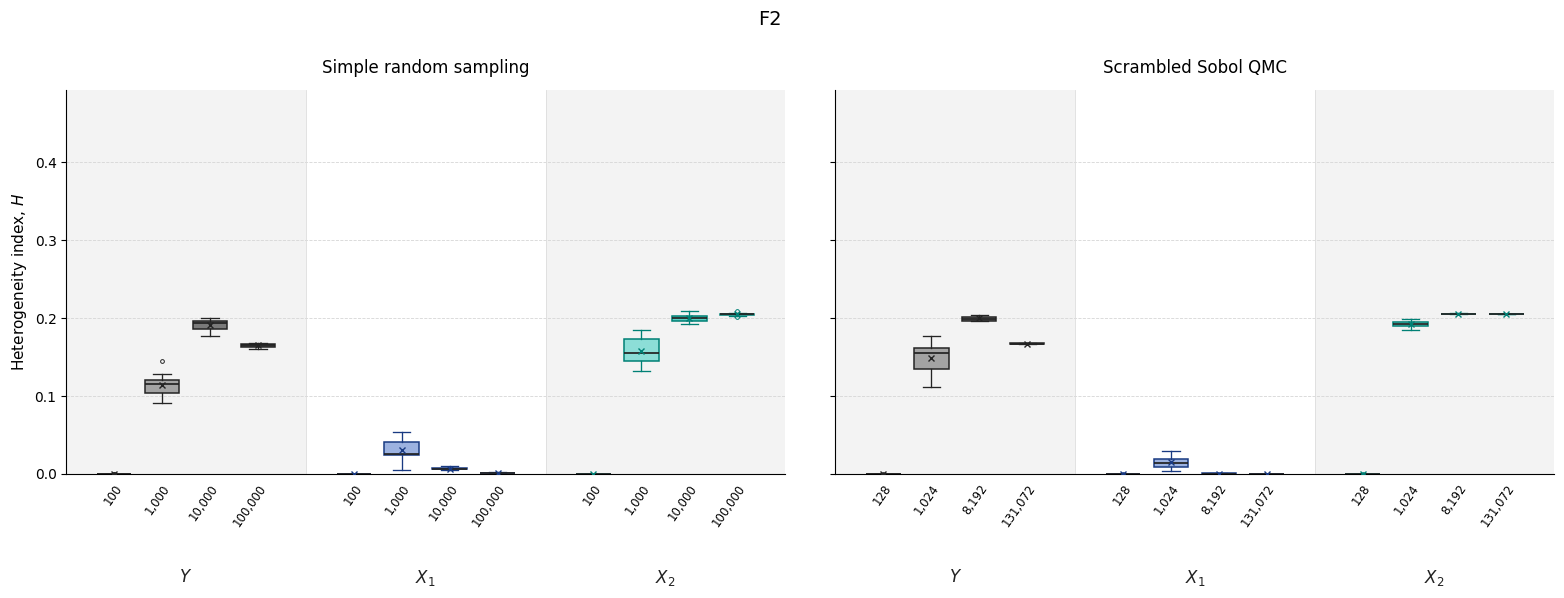}
    \captionof{figure}{Sample-size stability of the heterogeneity index $H$ for F2 function under simple random sampling and scrambled Sobol' QMC sampling.}
    \label{fig:b3}
\end{center}

\begin{center}
    \includegraphics[width=0.95\textwidth]{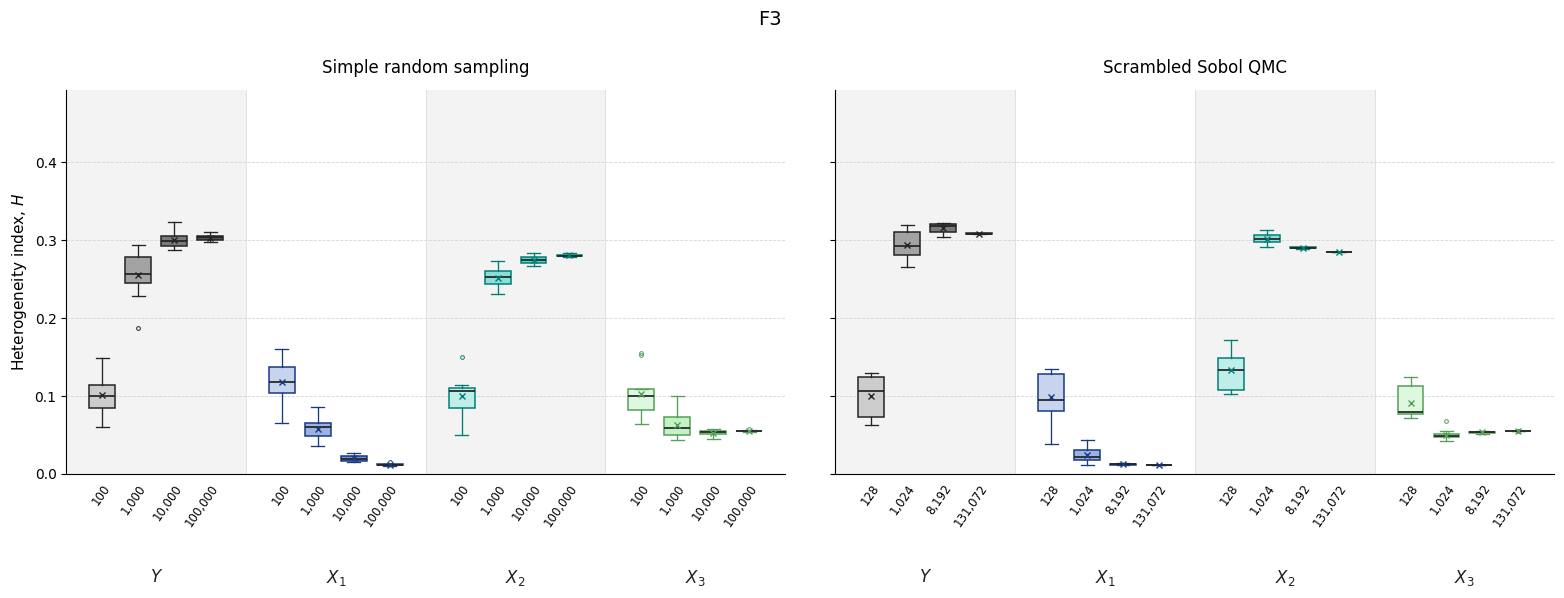}
    \captionof{figure}{Sample-size stability of the heterogeneity index $H$ for F3 function under simple random sampling and scrambled Sobol' QMC sampling.}
    \label{fig:b4}
\end{center}

\begin{center}
    \includegraphics[width=0.95\textwidth]{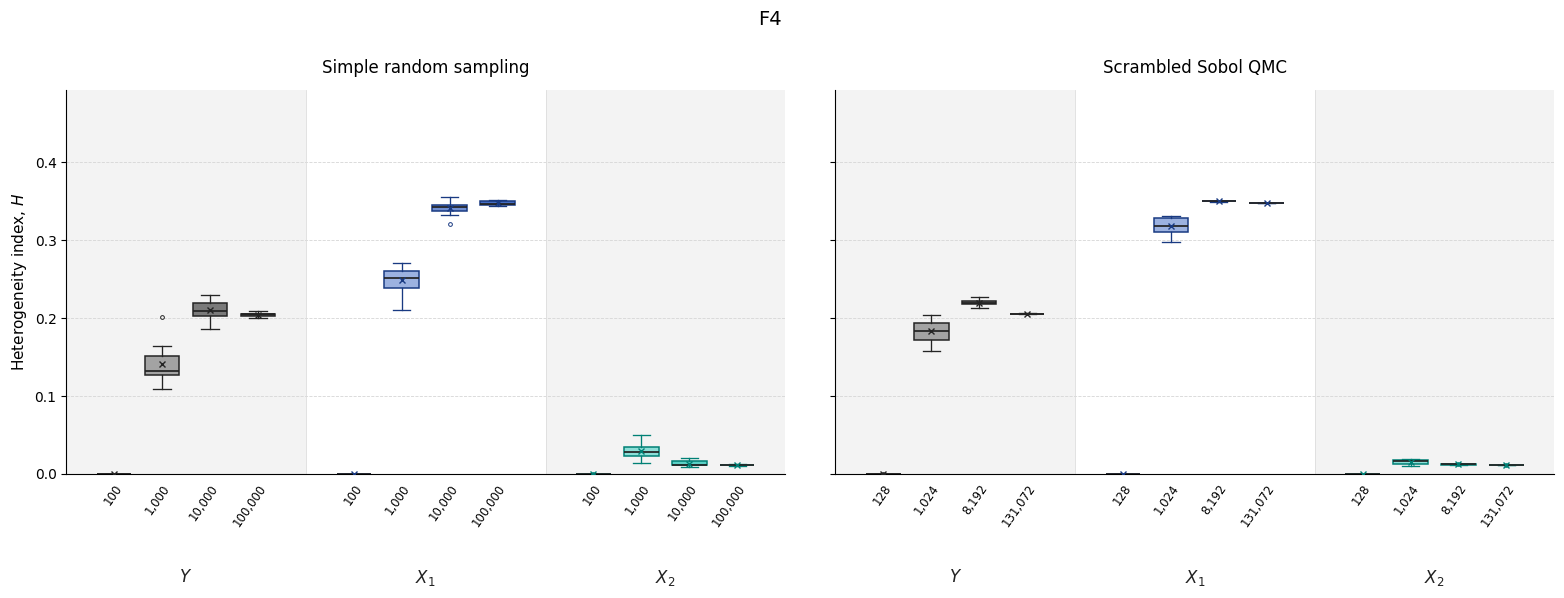}
    \captionof{figure}{Sample-size stability of the heterogeneity index $H$ for F4 function under simple random sampling and scrambled Sobol' QMC sampling.}
    \label{fig:b5}
\end{center}

\begin{center}
    \includegraphics[width=0.95\textwidth]{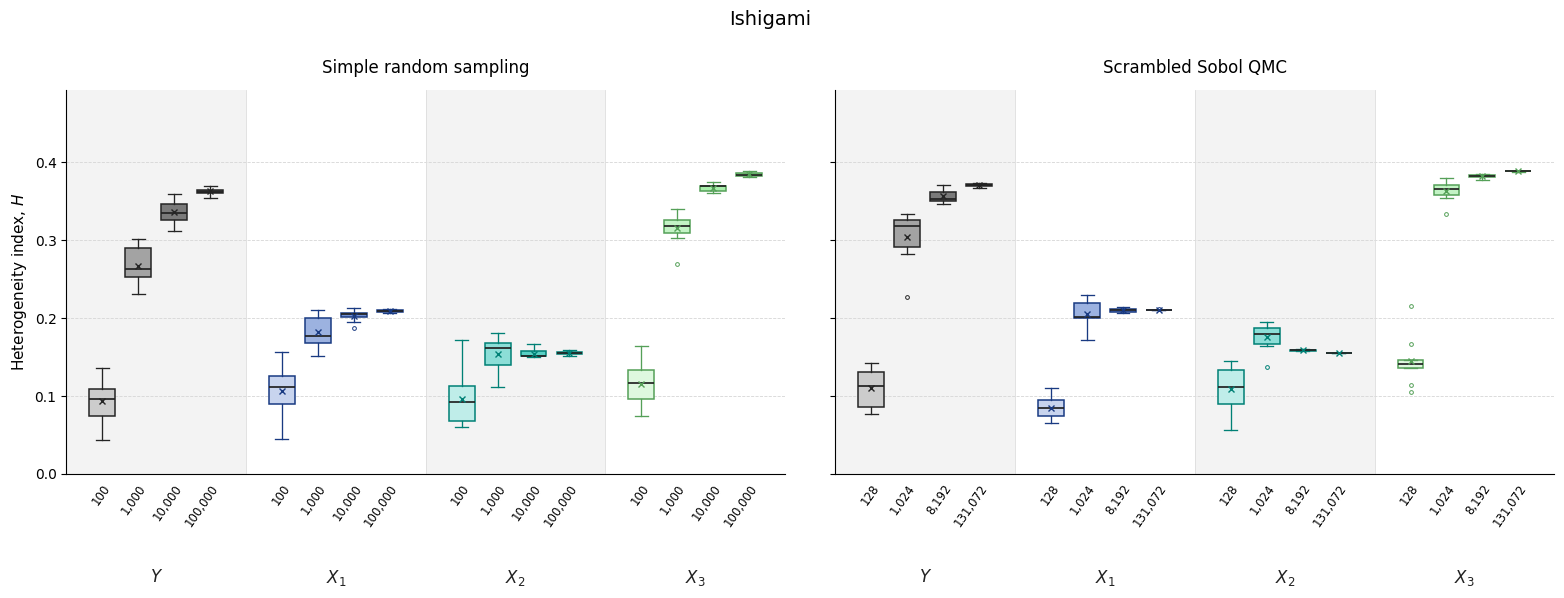}
    \captionof{figure}{Sample-size stability of the heterogeneity index $H$ for the Ishigami function under simple random sampling and scrambled Sobol' QMC sampling.}
    \label{fig:b6}
\end{center}

\begin{center}
    \includegraphics[width=0.95\textwidth]{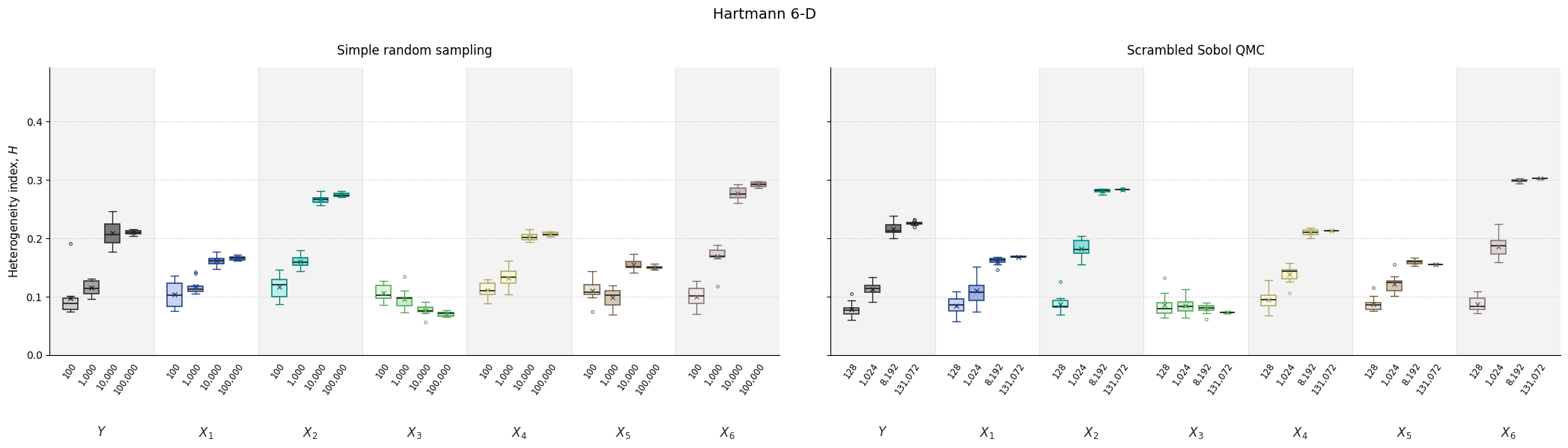}
    \captionof{figure}{Sample-size stability of the heterogeneity index $H$ for the Hartmann 6-D function under simple random sampling and scrambled Sobol' QMC sampling.}
    \label{fig:b7}
\end{center}

\begin{center}
    \includegraphics[width=0.95\textwidth]{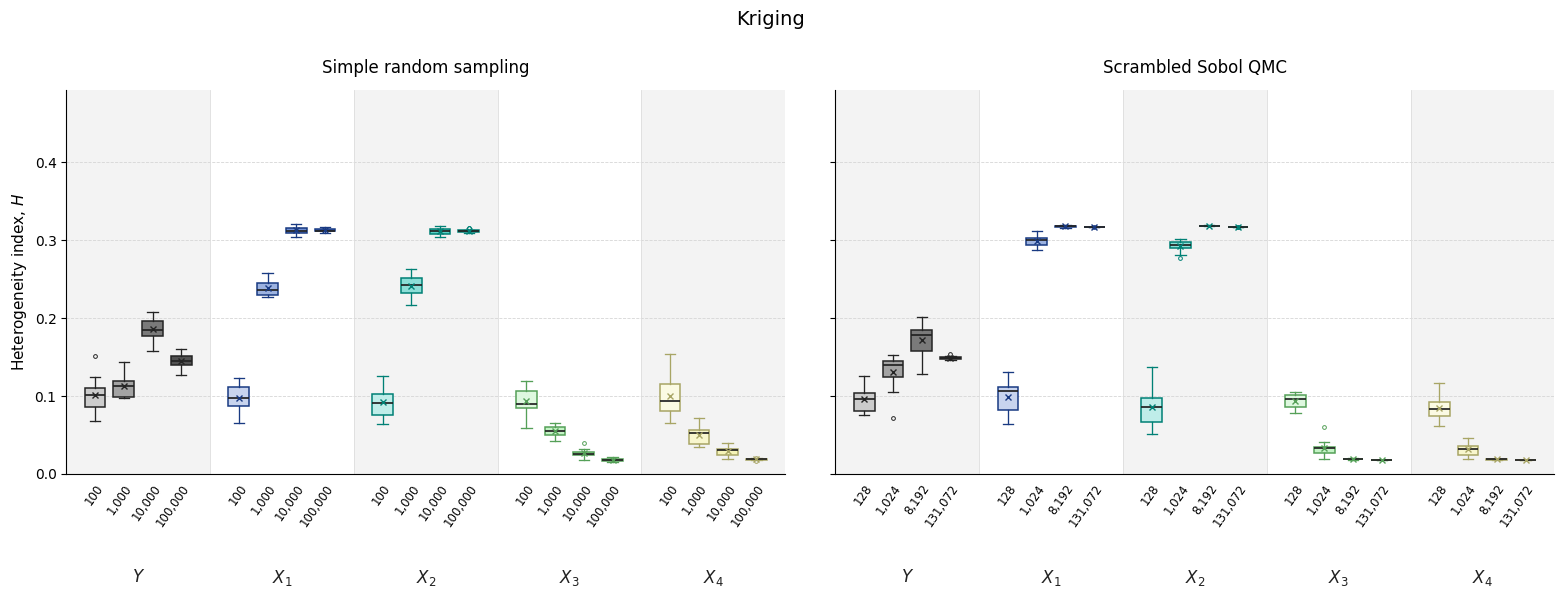}
    \captionof{figure}{Sample-size stability of the heterogeneity index $H$ for the Kriging function under simple random sampling and scrambled Sobol' QMC sampling.}
    \label{fig:b8}
\end{center}

\begin{center}
    \includegraphics[width=0.95\textwidth]{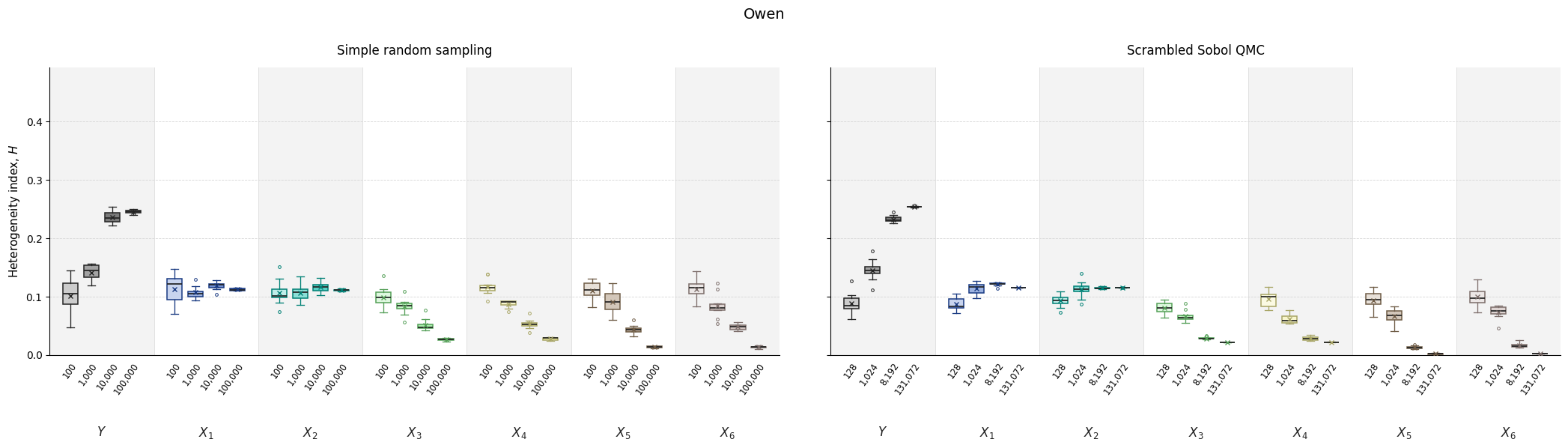}
    \captionof{figure}{Sample-size stability of the heterogeneity index $H$ for the Owen function under simple random sampling and scrambled Sobol' QMC sampling.}
    \label{fig:b9}
\end{center}

\end{document}